\pdfoutput=1
\documentclass[iop,apj]{emulateapj}
\usepackage{graphicx,epsfig}
\usepackage{rotate}
\usepackage{amsmath, amsthm, amssymb}
\usepackage{times}
\usepackage{soul}
\usepackage{natbib}
\usepackage[plainpages=false, colorlinks=true, anchorcolor=blue,
linkcolor=blue, citecolor=blue, bookmarks=false]
{hyperref}

\newcommand{\bq}{\begin{equation}}
\newcommand{\eq}{\end{equation}}

\def\gtsim{\lower.5ex\hbox{$\buildrel > \over\sim$}}
\def\ltsim{\lower.5ex\hbox{$\buildrel < \over\sim$}}

\def\sun{\mbox{$_\odot$}}
\def\apjl{ApJL}

\def\apjs{ApJS}

\def\aap{A\&A}

\shorttitle{Rotating PISN radiative transfer}
\shortauthors{Chatzopoulos, van Rossum, Wheeler, Whalen, Smidt, Wiggins}
\begin{document}
\title
{EMISSION FROM PAIR-INSTABILITY SUPERNOVAE WITH ROTATION}
\author{E. Chatzopoulos\altaffilmark{1,7}, van Rossum, Daniel R.\altaffilmark{1}, 
Wheeler, J. Craig \altaffilmark{2},
Daniel J. Whalen\altaffilmark{3}, Joseph Smidt\altaffilmark{4} \&
Brandon Wiggins\altaffilmark{5,6}}
\email{manolis@flash.uchicago.edu}
\altaffiltext{1}{Department of Astronomy \& Astrophysics, Flash Center for Computational
Science, University of Chicago, Chicago, IL, 60637, USA}
\altaffiltext{2}{Department of Astronomy, University of Texas at Austin, Austin, TX, 78712, USA}
\altaffiltext{3}{Universit\"{a}t Heidelberg, Zentrum f\"{u}r Astronomie, Institut f\"{u}r Theoretische Astrophysik, Albert-Ueberle-Str. 2, 69120 Heidelberg, Germany}
\altaffiltext{4}{T-2, Los Alamos National Laboratory, Los Alamos, NM 87545, USA}
\altaffiltext{5}{CCS-2, Los Alamos National Laboratory, Los Alamos, NM 87545, USA}
\altaffiltext{6}{Department of Physics and Astronomy, Brigham Young University, Provo, UT 84602}
\altaffiltext{7}{Enrico Fermi Fellow}

\begin{abstract}

Pair Instability Supernovae have been suggested as candidates for some
Super Luminous Supernovae, such as SN~2007bi,
and as one of the dominant types of explosion occurring in the early Universe from massive, zero-metallicity 
Population III stars. The progenitors of such events can be rapidly rotating, therefore exhibiting different
evolutionary properties due to the effects of rotationally-induced mixing and mass--loss.
Proper identification of such events requires rigorous radiation hydrodynamics and radiative transfer
calculations that capture not only the behavior of the light curve but also the spectral
evolution of these events. We present radiation hydrodynamics and radiation transport
calculations for 90-300~$M_{\odot}$ rotating pair-instability supernovae covering both the shock break-out
and late light curve phases. We also investigate cases of different initial metallicity and rotation 
rate to determine the impact of these parameters on the detailed
spectral characteristics of these events.
In agreement with recent results on non-rotating pair instability
supernovae, we find that for a range of progenitor masses and rotation
rates these events have intrinsically red colors in contradiction with observations of super-luminous supernovae.
The spectroscopic properties of rotating pair instability supernovae are similar to those of non-rotating events
with stripped hydrogen and helium envelopes. 
We find that the progenitor metallicity and rotation rate properties are erased after the explosion and cannot
be identified in the resulting model spectra. It is the combined effects of pre-supernova mass--loss and
the basic properties of the supernova ejecta such as mass, temperature and velocity that have the
most direct impact in the model spectra of pair instability supernovae.
\end{abstract}

\keywords{physical data and processes: radiative transfer --- stars: rotation --- stars: massive --- supernovae: general, supernovae: individual (pair-instability)}

\vskip 0.57 in

\section{INTRODUCTION}\label{intro}

The importance of understanding the supernova (SN) explosions of very massive 
stars ($M_{\rm ZAMS}>$~50~$M_{\odot}$) is pertinent in both low and high redshift. 
Recent developments in transient astrophysics led to the discovery of a new class
of relatively nearby ($z \sim$~0 - 1.2) explosions with peak luminosities $\sim$10-100 greater than those of
all types of core-collapse supernova (CCSN) events and with
a striking degree of diversity in their emission properties, the superluminous supernova (SLSN) events
(\citealt{2007ApJ...668L..99Q}; \citealt{2007ApJ...666.1116S}, see also \citealt{2012Sci...337..927G} for a review and more references).

Although the extreme luminosities of the majority of the observed SLSNe are attributable to other processes, such as
strong SN ejecta - circumstellar matter (CSM) interaction 
\citep{2007ApJ...671L..17S, 2011ApJ...729..143C, 2012ApJ...760..154C, 2012ApJ...747..118M, 2013MNRAS.428.1020M, 
2013ApJ...773...76C, 2014MNRAS.441..289B, 2014arXiv1405.1325N}
or efficiently thermalized proto-magnetar spin-down radiation 
\citep{2010ApJ...717..245K, 2010ApJ...719L.204W, 2012MNRAS.426L..76D, 2014ApJ...787..138L, 2014MNRAS.437..656M},
there are still a few SLSN light curves (LCs) that seem to be well fit by explosions of very massive progenitor stars 
($M_{\rm ZAMS} >$~130~$M_{\odot}$) powered by the radioactive decay of large amounts of $^{56}$Ni ($M_{\rm Ni}>$~1-10~$M_{\odot}$,
\citealt{2002ApJ...567..532H, 2007Natur.450..390W})
with the most well-studied case being SN~2007bi 
\citep{2009Natur.462..624G, 2012Sci...337..927G, 2011ApJ...734..102K, 2014A&A...565A..70K, 2014A&A...566A.146K}.

These pair-instability supernovae (PISNe; 
\citealp{1967ApJ...148..803R}; \citealp{1967PhRvL..18..379B}; \citealp{1967ApJ...150..131R}; \citealp{1983A&A...119...61O};
\citealp{2002ApJ...567..532H}; \citealp{2012ApJ...748...42C}; \citealp{2013ApJ...776..129C}; \citealp{2013MNRAS.433.1114Y}; 
\citealt{2014arXiv1402.5960C,2014arXiv1402.4134C})
occur when the helium cores of very massive stars enter a regime of high temperature ($\gtsim$~10$^{9}$~K)
and relatively low density ($\sim$~10$^{3}$-10$^{6}$~g~cm$^{-3}$) allowing rapid production of electron-positron (e$^{-}$-e$^{+}$) pairs
($M_{\rm ZAMS} \simeq$~130-260~$M_{\odot}$, $M_{\rm CO} \simeq$~60-140~$M_{\odot}$ for non-rotating progenitors
at ZAMS, where $M_{\rm CO}$ is the mass of the He core; \citealt{2002ApJ...567..532H, 2007Natur.450..390W}). 
Note that we use the ``CO'' notation since 75\% of the He core is, in reality, C and O.
The rapid electron-positron pair production leads to 
the reduction of radiation pressure in the core and the decrease of the adiabatic index, $\Gamma_{ad}$, below 4/3
resulting in core contraction and, consequently explosive 
nuclear burning and the production of a strong SN shock that totally disrupts the star. 

A few very massive stars (VMS) that seem to fit the criteria to be PISN progenitors
have been discovered \citep{2010MNRAS.408..731C}. At near to solar metallicities, models predict that
the effects of radiatively-driven mass--loss are so strong that they can drive
the mass of the star below the range required for PISNe (\citealt{2007A&A...475L..19L}; but
see \citealt{2014A&A...565A..70K} for an alternative view). In fact, the measured metallicities of
most SLSN hosts are always larger than $\sim$~0.1~$Z_{\odot}$ \citep{2013ApJ...773...12S, 2014ApJ...787..138L}.

The prevalence of PISNe is therefore expected to be higher in very low to zero metallicity environments
characteristic of the conditions in the early Universe. The Population III stars are
found to be quite massive in primordial star formation simulations,
well in the PISN regime \citep{1998ApJ...508..518A, 2002ApJ...564...23B, 2004ARA&A..42...79B, 2012MNRAS.422..290S, 2013MNRAS.431.1470S}. 
Pending the deployment of future missions such as the {\it James Webb Space Telescope} ({\it JWST})
and the {\it Wide Field Infrared Survey Telescope} ({\it WFIRST}), it has been broadly proposed 
to look for the primordial PISNe explosions resulting from the first stars 
\citep{2005ApJ...633.1031S, 2009Natur.459...49B, 2011ApJ...728..129J, 2012MNRAS.422.2701P, 2012ApJ...755...72H, 2013ApJ...762L...6W, 2013ApJ...777..110W,
2013ASSL..396..103G, 2013AcPol..53..573W, 2014MNRAS.442.1640D}.

Since PISNe can be linked to both SLSNe and primordial Pop III SNe, it is critical
to understand their radiative properties across the relevant parameter space.
So far the corresponding efforts have focused on non-rotating PISN
progenitors on the high end of the mass range. Gray flux limited
diffusion (FLD) radiation hydrodynamics model LCs have been presented 
in the context of the detectability of primordial PISNe 
\citep{2005ApJ...633.1031S, 2013ApJ...762L...6W, 2013ApJ...777..110W} but also
to compare to the observed LCs of some SLSNe such as SN~2006gy \citep{2007Natur.450..390W}
and SN~2007bi and PTF10nmn \citep{2009Natur.462..624G, 2011ApJ...734..102K, 2014A&A...565A..70K, 2014A&A...566A.146K}.

A rigorous, self-consistent comparison with observed SN spectra requires
accurate modeling of the observed spectral evolution and not just
the bolometric and broad band LCs. \citet{2013MNRAS.428.3227D} demonstrated this fact by
non local thermal equillibrium (non--LTE)
radiative transfer calculations yielding model non-rotating PISN spectra as a function of time with
the {\it CMFGEN} code \citep{2012MNRAS.424..252H}. They found that PISNe
are intrinsically too red in color as compared to SLSN events like SN~2007bi,
and that their spectral evolution does not match that of any observed SN
so far. 

Recently, it has been shown that the inclusion of the effects of rotation
in the evolution of PISN progenitors changes some key characteristics
of these events 
\citep{2012ApJ...760..154C, 2012ApJ...748...42C, 2012A&A...542A.113Y, 2013ApJ...776..129C}. 
More specifically, rotationally--induced mixing can shift the mass limit to encounter PISN to lower
values (85~$M_{\odot}$~$<M_{\rm ZAMS}<$~190~$M_{\odot}$ at rotation
rates 50~\% that of the critical Keplerian value). In addition, the
combined effects of mixing and rotationally--induced mass--loss
often lead to bare CO cores (stripped totally of both H and He) 
prior to explosion and to differences
in mixing, $^{56}$Ni yields, and energetics, therefore changing
the interplay between energy input and SN ejecta opacity. This alone
may have an effect on the radiative properties of these events
as compared to non-rotating PISNe. 

Rapid stellar rotation
has been observed for some massive nearby stars \citep{2011ApJ...743L..22D}
and it is also found in simulations of primordial massive
star formation \citep{2011ApJ...737...75G, 2013MNRAS.431.1470S}.
To further complement the parameter space relevant to PISN explosions,
in this work we present radiation hydrodynamics
and LTE radiative transfer calculations of
PISNe resulting from rotating progenitors and compare our results
to the non-rotating case and to some observed SNe and 
the PISN candidate SN~2007bi.

The paper is organized as follows. In Section~\ref{Stellevol} we present the 1D stellar evolution
and hydrodynamics simulations of rotating PISN progenitors up to 
a time prior to shock break out (SBO). In Section~\ref{SBOemission} we discuss the shock break out
emission using results from radiation hydrodynamics simulations. In Section~\ref{RadTrans}
we present LTE LCs and spectra of rotating PISNe and compare
with results from non-rotating models \citep{2011ApJ...734..102K, 2013MNRAS.428.3227D} as well
as observations of SLSNe like SN~2007bi and other events. Lastly, 
in Section~\ref{Disc} we summarize and discuss our conclusions.

\section{STELLAR EVOLUTION AND HYDRODYNAMICS SIMULATIONS}\label{Stellevol}

In order to obtain time-dependent spectra and final LCs for rotating PISNe, we followed
a multi-step process. First we evolved a grid of progenitors
from the Zero Age Main Sequence (ZAMS) up to the point when the dynamical
instability is encountered using the stellar evolution code Modules for Experiments
in Stellar Astrophysics ({\it MESA}; \citealt{2011ApJS..192....3P, 2013ApJS..208....4P}). 
At this point, the structure of the star was becoming dynamical and the semi-hydrostatic stellar
evolution code reached its limitations.
For this reason, we conservatively mapped the {\it MESA} models to the 1D Adaptive
Mesh Refinement (AMR) grid of the multi-physics hydrodynamics code {\it FLASH}
\citep{2000ApJS..131..273F, 2009arXiv0903.4875D} and followed the evolution of the explosion
up to the epoch before SN SBO. 
Since we needed to capture the full radiation hydrodynamics nature of SN SBO involving
the separation of matter and radiation temperatures, the next step involved re-mapping of
the blast profiles to the radiation
hydrodynamics code {\it RAGE} \citep{2008CS&D....1a5005G, 2013ApJS..204...16F}.
The {\it RAGE} profiles were post-processed by the Los Alamos National Lab (LANL)
{\it SPECTRUM} \citep{2013ApJS..204...16F} code yielding SBO LCs and spectra.
Finally, once the SN ejecta were expanding homologously the post-SBO profiles were mapped in
the Langrangian grid of the radiative transfer code {\it PHOENIX} 
\citep{1992JQSRT..47..433H, 1999JCoAM.109...41H, 2004A&A...417..317H, 2012ApJ...756...31V}
that provided us with  time-dependent spectra and LCs of rotating PISNe. We only calculated
SBO LCs for 10 zero metallicity rotating models
because PISN models of higher metallicity and zero rotation have already been
investigated \citep{2011ApJ...734..102K,2013MNRAS.428.3227D}.

\subsection{{\it MESA pre-supernova evolution.}}\label{MESApreSN}

We modeled the evolution of 10 PISN progenitor stars
with zero initial ZAMS metallicity, rotation at 50\% that of the critical Keplerian value
at the equator and in the mass range 90-140~$M_{\odot}$ using
{\it MESA} version 5596. This mass range was motivated by the finding
of \citet{2012ApJ...748...42C} and \citet{2012A&A...542A.113Y}
that full-fledged zero metallicity PISNe 
that rotate with $\Omega/\Omega_{\rm c,ZAMS}=$~0.5 
occur for ZAMS masses $>$~85~$M_{\odot}$.
In the above expression, $\Omega$ and $\Omega_{\rm c,ZAMS}$ 
are the angular velocity at the equator and the critical value of angular velocity at ZAMS,
respectively. 

The zero metallicity models were considered because of their relevance
to primordial star formation and the proposals to look for the PISN explosions
from the first massive stars 
\citep{2005ApJ...633.1031S, 2013ApJ...762L...6W, 2013ApJ...777..110W}.
For that same reason, we adopted an initial rotation rate of $\Omega/\Omega_{c,ZAMS}=$~0.5
that seems to be a characteristic value found in early Universe star formation simulations
\citep{2011ApJ...737...75G, 2013MNRAS.431.1470S}. 

In order to properly assess the parameter space, but also to study models relevant
to nearby recently discovered PISN candidates such as SN~2007bi, we calculated
two rotating PISN models at  higher metallicities (260~$M_{\odot}$ at 0.05~$Z_{\odot}$ and
300~$M_{\odot}$ at 0.1~$Z_{\odot}$ at the ZAMS both with $\Omega/\Omega_{\rm c,ZAMS}=$~0.5). 
These large ZAMS masses are characteristic of the most massive stars observed \citep{2010MNRAS.408..731C}.
Very large ZAMS mass is required in order to counter the strong effects of radiatively-driven mass-loss at these high metallicities
allowing the formation of final cores with masses in the PISN regime.
To determine whether the presence of metals in otherwise identical PISNe can be detected in 
model spectra, we artificially added metal content in the 140~$M_{\odot}$ model assuming
$Z =$~0.1~$Z_{\odot}$ prior to mapping to {\it PHOENIX} and re-normalized all the abundances (``pP'' model). 
In addition, to determine whether the presence of the same metal content in the 140~$M_{\odot}$ 
model has an effect on the explosive PISN nucleosynthesis we also added the same metal content (0.1~$Z_{\odot}$)
prior to mapping to {\it FLASH} (``pF'' model). Although the models with artificially added metallicity (``pP'' and ``pF'') are
not entirely self-consistent since the presence of metals will itself change the pre-SN evolution, we use
these models to gain intuition on the magnitude of the effect.  

Finally, to probe spectroscopic differences that might arise due to varying degrees of pre-PISN
rotation rate, we calculated two zero metallicity non-rotating models (with $M_{\rm ZAMS} =$~200
and 260~$M_{\odot}$) and a 140~$M_{\odot}$ model with $\Omega/\Omega_{\rm c,ZAMS}=$~0.5 but without the effects
of mixing and angular momentum transport due to the magnetic fields by turning
off the Spruit-Taylor (ST) mechanism (140~$M_{\odot}$ ``no-ST'' model). The absence of magnetic viscosity in this model
led to a much faster rotating pre-PISN core ($\Omega/\Omega_{\rm c} \simeq$~0.3).
As a result we calculated a total of four 140~$M_{\odot}$ models spanning the rotation and metallicity
parameter space: the original model with the effects of ST included, one without the effects of ST and two
with artifically added metallicity (0.1~$Z_{\odot}$) prior to doing the PISN nucleosynthesis in {\it FLASH}
and prior to mapping to {\it PHOENIX} to calculate the radiative transfer properties. 
For details on the {\it MESA} parameters used (equation of state, nuclear reaction network,
resolution, treatment of mass--loss and rotation) refer to \citet{2012ApJ...748...42C}. The basic
properties of the PISN models we examine in this work are listed in Table~\ref{T1}.

\begin{figure}
\begin{center}
\includegraphics[angle=-90,width=9cm,trim=0.8in 0.25in 0.5in 0.15in,clip]{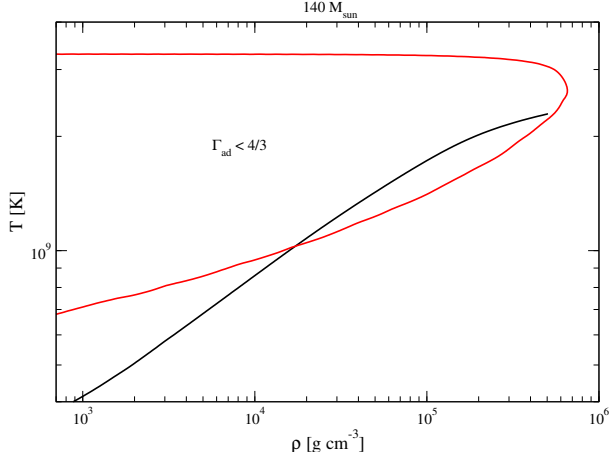}
\caption{The black curve shows the nitial {\it MESA} progenitor $\rho$-$T$ profile 
for the 140~$M_{\odot}$ model prior to mapping to the grid of {\it FLASH}. 
The thick red curve denotes the region where $\Gamma_{ad}<$~4/3 
due to the effects of e$^{-}$- e$^{+}$ pair production.\label{Fig:140sm_struc}}
\end{center}
\end{figure}

\begin{figure}
\begin{center}
\includegraphics[angle=-90,width=9cm,trim=0.8in 2.8in 0.5in 2.1in,clip]{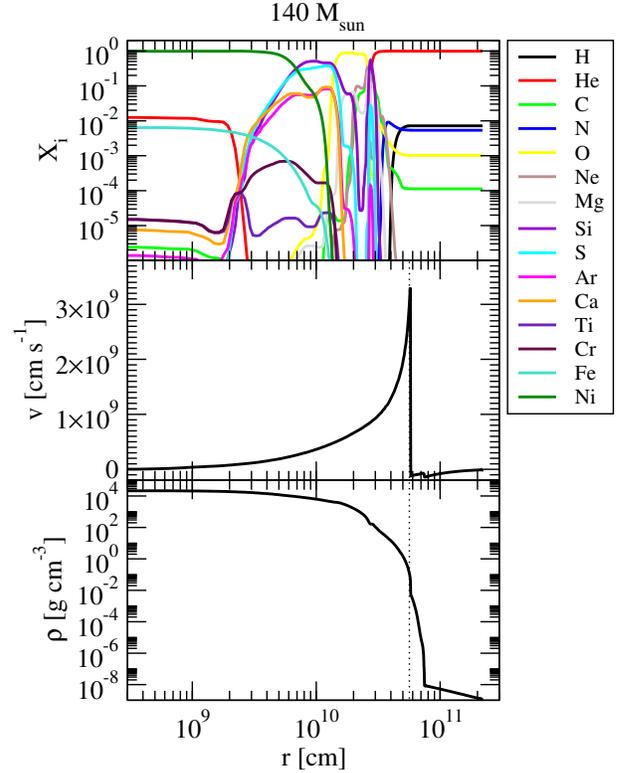}
\caption{1D {\it FLASH} radial profiles of composition (upper panel), 
velocity (middle panel) and density (lower panel)
for the 140~$M_{\odot}$ model prior to mapping to the grid of {\it RAGE} just before SN shock break-out. 
The dotted vertical lines correspond to the location of the SN shock.\label{Fig:140sm_flash}}
\end{center}
\end{figure}

The effects of rotation and magnetic fields via the ST mechanism 
\citep{1999A&A...349..189S, 2002A&A...381..923S} on the evolution of massive stars have been investigated 
thoroughly in the past with the use of different stellar evolution codes 
\citep{2000ApJ...528..368H, 2005A&A...443..643Y, 2011A&A...530A.115B, 2011A&A...530A.116B, 
2008A&A...489..685E, 2012A&A...537A.146E, 2012ApJ...760..154C, 2012A&A...542A.113Y}, also
see \citep{2011arXiv1109.6171M} for a review. Enhanced chemical mixing, primarily
due to meridional circulation and the ST mechanism, can lead to the formation
of more massive CO cores for a lower ZAMS mass than required in the case of zero rotation. 
Furthermore, diffusion of angular momentum can lead to additional, 
mechanically-induced mass--loss \citep{2000ApJ...528..368H} that in turn can, under
some circumstances, result in the stripping of both the H and He envelopes
from the SN progenitor star leaving a bare CO core remnant. 

This is the case for most of the models we consider in this work. In particular, we stopped
the {\it MESA} calculations when a significant portion of the pre-PISN CO cores
was in the pair-instability regime of $\Gamma_{\rm ad}<$~4/3 and the model
was becoming dynamical. In general this was the phase of core carbon
exhaustion, in agreement with \citep{2013MNRAS.428.3227D}.
Figure~\ref{Fig:140sm_struc} shows the temperature-density ($\rho$-$T$)
structure of the 140~$M_{\odot}$ model when the {\it MESA} run was interrupted.
Note that a significant portion of the model lies within the pair-instability regime
marked by the thick red curve. 

For all of the models considered here
the final, pre--SN mass of the CO core remnants ($M_{\rm CO}$) was 
in the range of $\sim$~60-112~$M_{\odot}$ and the final radius, $R_{\rm f}$, in
the range $4 \times 10^{10}$-$4 \times 10^{13}$~cm. 
The rotating models consintently made more compact PISN progenitors due to
the effects rotationally-enhanced mass-loss.
Due to angular momentum transport during the main evolution, the final, pre--SN
rotation rates of the CO cores were somewhat modest 
($\Omega/\Omega_{\rm c} \simeq$~0.02-0.06) with the exception of the 140~$M_{\odot}$ no-ST model.
As such, most of these pre--SN models have somewhat different structural properties
than other PISN models considered so far in the literature
for radiative transfer studies \citep{2009Natur.462..624G, 2011ApJ...734..102K, 2013MNRAS.428.3227D}.
They are compact, well chemically-mixed 
slowly rotating CO stars with the H envelope removed in all cases and with a low, even zero
in a few cases, He envelope mass.

\setcounter{table}{0}
\begin{deluxetable*}{ccccccccc}
\tablewidth{0pt}
\tablecaption{Properties of PISN progenitor models.}
\tablehead{
\colhead {$M_{\rm ZAMS}$~($M_{\odot}$)} &
\colhead{$Z_{\rm ZAMS}$~($Z_{\odot}$)} &
\colhead {$R_{\rm f}$~($10^{10}$~cm)} &
\colhead {$M_{\rm f}$~($M_{\odot}$)} &
\colhead {$M_{\rm CO}^{\dagger}$~($M_{\odot}$)} &
\colhead {$M_{\rm Ni}$~($M_{\odot}$)} &
\colhead {$E_{\rm SN}-E_{\rm b,f}$$^{*}$~($10^{52}$~erg)} &
\colhead {$M_{\rm bol,peak}$~(mag)} &
\\}
\startdata
90      & 0.0  &  3.925    & 60.51 &  59.31  &   0.138    &  0.98  &  -16.08    \\  
95      & 0.0  &  5.598    & 68.97 &  63.48  &   0.327    &  1.07  &  -16.93    \\  
105    & 0.0  &  4.730    & 71.12 &  69.10  &   0.853    &   2.84 &   -17.96   \\   
110    & 0.0  &  5.707    & 74.93 &  70.35  &   1.141    &   3.94 &   -18.15   \\   
115    & 0.0  &  8.266    & 76.73 &  70.48  &   1.070    &   3.12 &   -18.05   \\		 
120    & 0.0  &  6.565    & 79.03 &  72.57  &   1.565    &   4.28 &   -18.52   \\   
125    & 0.0  &  8.034    & 83.60 &  76.80  &   3.255    &   5.04 &   -19.30   \\   
130    & 0.0  & 10.616   & 85.64 &  77.69  &   3.870    &    5.23 &  -19.47   \\    
135    & 0.0  & 18.324   & 90.15 &  79.76  &   4.521    &    6.16 &  -19.66   \\    
140    & 0.0  &  7.649    & 91.30 &  83.71  &   7.297    &   7.95 &   -20.24   \\  
140pF    & 0.1  &  7.649    & 91.30 &  85.60   &   6.453    &  7.62 &  -20.24  \\
140pP    & 0.1  &  7.649    & 91.30  & 83.71   &   7.297    &  7.95 &  -20.24   \\
140no-ST    & 0.0  & 22.341   & 91.82  & 85.17   &   2.489    &  3.08 & -19.20   \\
200$^{a}$    & 0.0  &3772.167& 111.28 &  82.48  &   0.165    &  1.55 & -16.00    \\
260$^{a}$    & 0.0  &1398.403& 139.43 &  95.11  &   9.858    &  4.94 & -20.50    \\  
260    & 0.05& 7.299   & 116.85  &  112.02  &   22.22    &   6.64  & -21.50   \\
300    & 0.1  & 9.322   & 111.02 &  111.02  &   12.70    &   4.78  & -20.92   \\
\enddata 
\tablecomments{$^{\dagger}$ $M_{\rm CO}$ is the mass of the carbon-oxygen core defined as the total mass
within the radius where $X_{\rm C}+X_{\rm O} >$~0.5. 
$^{*}$ The nuclear energy released by the SN explosion, $E_{\rm SN}$, minus the pre-SN binding energy of the
progenitor, $E_{\rm b,f}$.
$^{a}$ These models were evolved without rotation.
\label{T1}}
\end{deluxetable*}

\subsection{{\it 1-D FLASH hydrodynamics.}}\label{Sec:1dFLASH}

All the pre--SN 1D profiles from {\it MESA} were mapped to the 1D
AMR Eulerian grid of {\it FLASH}-4.0 \citep{2000ApJS..131..273F, 2009arXiv0903.4875D}. 
The same basic physics units (equation of state,
nuclear reaction network) are used in {\it FLASH} 
as in the {\it MESA} stellar evolution calculations with the
exception that we did not map the pre--SN rotational profile for the hydrodynamics
simulation. As \citet{2013ApJ...776..129C} have shown, when the
magnetic viscosity and angular momentum transport due to the ST dynamo action are
included in the stellar evolution calculation, the pre--SN rotation is slow 
and does not affect the final hydrodynamics. It is the pre--SN mixing changing the 
structure of the progenitor star and the composition
gradient effects that primarily determine the final dynamics and nucleosynthetic properties
of the explosion. Also, the oblateness of the star due to rotation will be negligible for the
moderate pre--SN rotational velocities of our suite of models.

We follow the collapse, explosion and nucleosynthesis in {\it FLASH} up until prior to
SN SBO, that we take to be the phase when the SN shock is within
a hundred to a few thousand optical depths from the surface
of the star, still in the optically thick regime. We recall that $\tau_{\rm SBO} = c/v_{\rm sh}$ where
$\tau_{\rm SBO}$ is the SBO optical depth, $v_{\rm sh}$ the velocity of the SN shock (typically of
the order of a few 10$^{9}$~cm~s$^{-1}$) and $c$ the speed of light. 
Figure~\ref{Fig:140sm_flash} shows the composition (upper panel),
velocity (middle panel) and density (lower panel) for the pre-SBO stage when we halt the
{\it FLASH} simulation in the case of the 140~$M_{\odot}$ model. As illustrated in 
Figure~\ref{Fig:140sm_flash}, the basic structure of this model consists of $^{56}$Ni
located in the base of H-poor SN ejecta. Table~\ref{T1} details
the basic characteristics of all PISN explosion models. Our set of models produced
final $^{56}$Ni masses in the range 0.1-22~$M_{\odot}$ embedded in 
SN ejecta of 60-112~$M_{\odot}$. The $M_{\rm Ni}/M_{\rm f}$ ratios for our rotating 
90-140~$M_{\odot}$ and non-rotating 200~$M_{\odot}$ and 260~$M_{\odot}$ PISN models are consistent with those
of \citet{2013MNRAS.428.3227D} meaning that, modulo differences in SN ejecta opacity,
our models are expected to manifest as dimmer events (in terms of bolometric LC) that have rather red colors. 
We test this prediction in Section~\ref{RadTrans}.

\section{ROTATING PISN SHOCK BREAKOUT EMISSION}\label{SBOemission}

As an intermediate step, the 10 final rotating zero metallicity PISN {\it FLASH} profiles were mapped into the 1D AMR Eulerian grid of 
the radiation hydrodynamics code {\it RAGE} \citep{2008CS&D....1a5005G, 2013ApJS..204...16F} where
the emission during the SBO phase is modeled. We did not map the other models into {\it RAGE} because
higher metallicity and non-rotating PISN SBO properties have already been well studied in the literature
\citep{2013ApJ...762L...6W,2013ApJ...777..110W}.
The SBO LCs are calculated by post-processing the {\it RAGE} snapshots with the LANL 
code {\it SPECTRUM} \citep{2013ApJS..204...16F}. For details on the physics, resolution
and other parameters used in {\it RAGE} and {\it SPECTRUM} please see \citet{2013ApJ...762L...6W}.

The {\it RAGE} and {\it SPECTRUM} results on the $^{56}$Ni radioactive decay powered re-brightening seem
to conflict with earlier results and a code comparison effort is underway to determine the source
of this discrepancy \citep{2013ApJ...777..110W}. 
Possible sources include differences in opacities, resolution and the issue of
what fraction of the radioactive decay energy goes into increasing the kinetic energy of the inner SN ejecta versus 
pure radiative losses. Homologous LTE and non--LTE codes are engineered to predict behavior closer to the latter
while {\it RAGE} radiation hydrodynamics favors the former case. In this paper, we are only using {\it RAGE}
to model the SBO emission during the first few days of radiation hydrodynamical evolution while for the
later evolution (homologous expansion phase) we use the LTE treatment of the {\it PHOENIX}
code (Section~\ref{RadTrans}).

The SBO LCs have some dependence on the properties of the CSM that is superimposed
on the SN blast profiles in the {\it RAGE} grid. For the purposes of this study, the abrupt density gradient
between the edge of the star and the wind is bridged by an $r^{-20}$ profile to minimize effects
of numerical instabilities during the SBO phase. Then the wind profile follows an inverse square power-law
given explicitly by $\rho_{w} = \dot{M}/(4\pi v_{w} r^{2})$ where $\rho_{w}$ is the wind density, $\dot{M}$
the wind mass--loss rate and $v_{w}$ the wind velocity taken to be 1000~km~s$^{-1}$. For all models
we fix the density at the bottom of the wind profile to be $\sim 2 \times 10^{-18}$~g~cm$^{-3}$ so that
the CSM is optically thin.

\begin{figure}
\begin{center}
\includegraphics[angle=-90,width=9cm,trim=1.in 0.25in 0.5in 0.15in,clip]{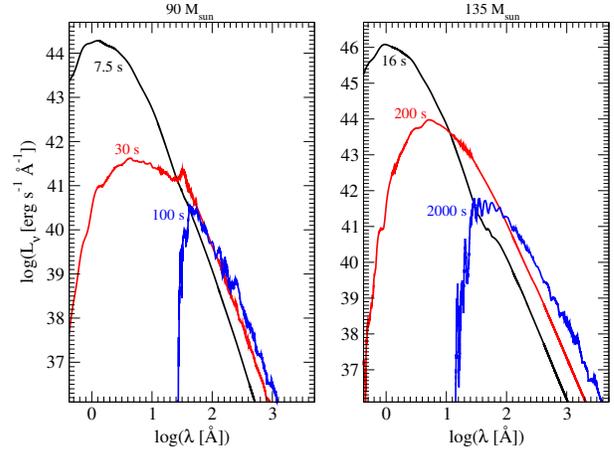}
\caption{Post-SBO SED evolution for the 90~$M_{\odot}$ (left panel)
and the 135~$M_{\odot}$ (right panel) model as calculated by {\it SPECTRUM}. 
The SED phases are shown on top of the corresponding curves.
The SBO emission for a bare CO rotating PISNe peaks in the hard X-rays (0.1-10~\AA~ or $\sim$~1-100~keV).\label{Fig:SBO_spec}}
\end{center}
\end{figure}

Using {\it SPECTRUM}, the spectral energy distributions (SEDs) during the first few days after SBO are
calculated and then integrated yielding the final bolometric LCs. In Figure~\ref{Fig:SBO_spec} the first few minutes of
post-SBO SED evolution for the 90~$M_{\odot}$ and the 135~$M_{\odot}$ models are shown. 
The SBO spectra for all of our models exhibit a characteristic ``expanding fireball'' behavior 
that is well described by a cooling black body continuum with some spectral lines superimposed, especially during
the later phases. The peak of the emission during SBO is in the hard X-ray part of the spectrum (0.1-10~\AA~
or $\sim$~1-100~keV) at significantly shorter wavelengths than SBO emission from red supergiant (RSG)
and blue supergiant (BSG) PISN progenitors \citep{2011ApJ...734..102K}. 

\begin{figure}
\begin{center}
\includegraphics[angle=-90,width=9cm,trim=1.in 0.25in 0.5in 0.15in,clip]{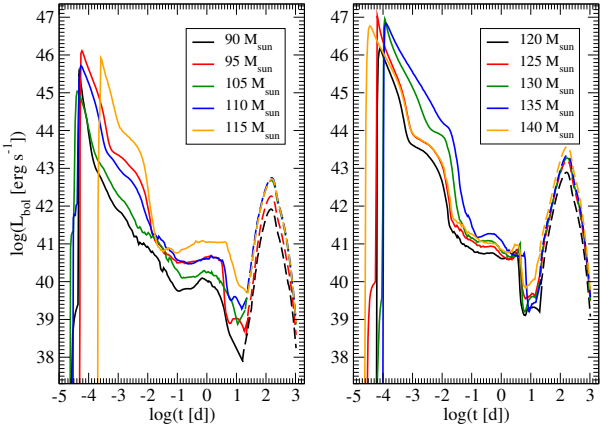}
\caption{{\it RAGE/SPECTRUM} SBO (solid curves) and {\it PHOENIX}  late-time (dashed curves) 
LCs for the grid of rotating PISNe models studied here. {\it Left Panel}: 90-115~$M_{\odot}$. 
{\it Right Panel}: 120-140~$M_{\odot}$.\label{Fig:Full_LCs}}
\end{center}
\end{figure}

Figure~\ref{Fig:Full_LCs} shows the resulting SBO bolometric LCs (solid curves) for all of the models discussed
in this paper. The late-time {\it PHOENIX}  LCs are also shown (dashed curves) for completeness but discussed
in detail in Section~\ref{RadTrans}. 
The peak SBO luminosities for all models are in the range 10$^{45}$-10$^{46}$~erg~s$^{-1}$
and are consistent with the \citet{2011ApJ...734..102K} results despite the different progenitor properties. This is mainly
because although our post-SBO temperatures and velocities 
are higher and the emission peaks at lower wavelengths, the radii of our
progenitor models are smaller than both the RSG and BSG series presented in \citet{2011ApJ...734..102K}. The rotating
PISN SBO bursts are also found to be close in terms of duration with those for BSG progenitors in \citet{2011ApJ...734..102K}.
After the end of the SBO bursts (0.01-0.1~days),
the luminosities are mainly driven by hydrodynamical effects (reverse and forward shock heating)
before the much longer re-brightening phase due to the radioactive decays of $^{56}$Ni and $^{56}$Co starts. 

\section{TIME-DEPENDENT RADIATIVE TRANSFER}\label{RadTrans}

In order to calculate time-dependent model spectra for our rotating PISN models
we first converted the blast profiles from {\it FLASH} at 1~hr after SBO into the Langragian
grid using velocity as the independent variable. These profiles
are then input to the radiative transfer code {\it PHOENIX} 
\citep{1992JQSRT..47..433H, 1999JCoAM.109...41H, 2004A&A...417..317H, 2012ApJ...756...31V}, 
allowing us to retrieve model light curves, color and spectral evolution.

\begin{figure}
\begin{center}
\includegraphics[angle=-90,width=9cm,trim=1.in 0.25in 0.5in 0.15in,clip]{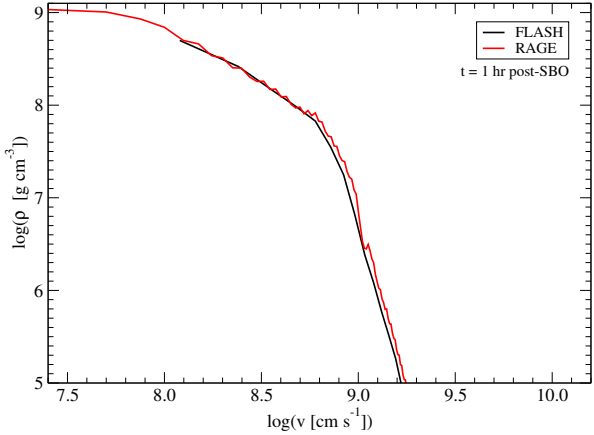}
\caption{Scaled density profile in Lagrangian coordinates at 1~hr after SBO in
{\it FLASH} (black curve) and {\it RAGE} (red curve). It can be seen that
the pure hydrodynamics and radiation hydrodynamics profiles are in good
agreement shortly after SBO as discussed in \S~\ref{RadTrans}.\label{Fig:1hr_comp}}
\end{center}
\end{figure}

At first, it might seem questionable that we elected to use the {\it FLASH} post-SBO
profiles instead of the {\it RAGE} ones that properly model the radiation hydrodynamic
effects during that early phase; however, once the SBO effects have ceased we find
that the density and mass fraction profiles calculated by the two codes are almost
identical up to $\sim$~1-4~days, depending on the model. To illustrate this we
present the Lagrangian density profiles for the 140~$M_{\odot}$ model 
at 1~hr in Figure~\ref{Fig:1hr_comp}
for both {\it FLASH} (black curve) and {\it RAGE} (red curve). 
We have verified that the resulting final model
LCs and spectra are virtually invariable with respect to what input profile is
used at 1~hr after SBO ({\it FLASH} or {\it RAGE}).

{\it PHOENIX} solves the special relativistic radiative transfer
problem in spherical symmetry using the highly accurate short 
characteristics and and operator splitting methods \citep{1987JQSRT..38..325O}.
The time-evolution of the post-hydrodynamic, radiation-dominated phase
is followed using the Radiation 
Energy Balance method \citep{2012ApJ...756...31V}.
{\it PHOENIX} does not use the Sobolev approximation, diffusion approximations, or opacity binning.
The deterministic nature of the radiative transfer method enables
calculation of noise-free spectra, a 
feature which enables us to show detailed line identifications for the presented model spectra (see next paragraph).

\subsection{{\it Zero metallicity 90-140~$M_{\odot}$ PISNe.}}\label{Sec:LTEZ0}

\begin{figure}
\begin{center}
\includegraphics[angle=-90,width=9cm,trim=0.9in 0.25in 0.5in 0.15in,clip]{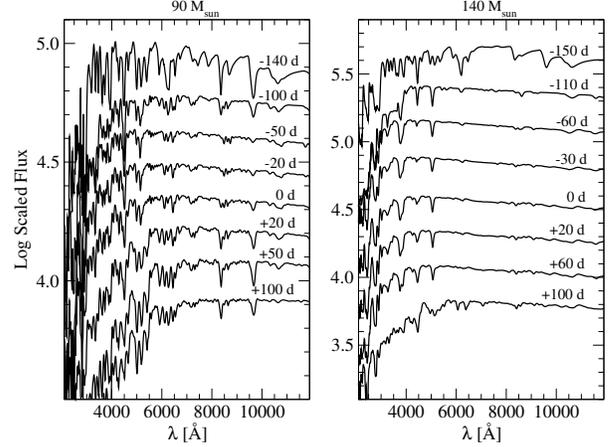}
\caption{{\it PHOENIX} spectral evolution for the 90~$M_{\odot}$ (left panel) and the 140~$M_{\odot}$ (right panel)
models. The phases are shown with respect to the time of $^{56}$Ni decay peak luminosity.\label{Fig:LTE_specevol}}
\end{center}
\end{figure}

\begin{figure}
\begin{center}
\includegraphics[angle=-90,width=9cm,trim=1.in 0.25in 0.5in 0.15in,clip]{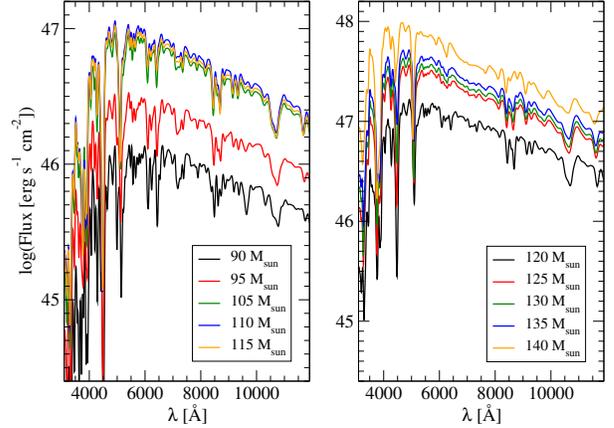}
\caption{{\it PHOENIX} spectra at peak luminosity for the zero metallicity rotating 90-140~$M_{\odot}$ PISNe. 
{\it Left Panel}: 90-115~$M_{\odot}$. {\it Right Panel}: 120-140~$M_{\odot}$.\label{Fig:Peak_LTEspec}}
\end{center}
\end{figure}

\begin{figure*}
\centerline{
 \includegraphics[width=.5\textwidth, clip]{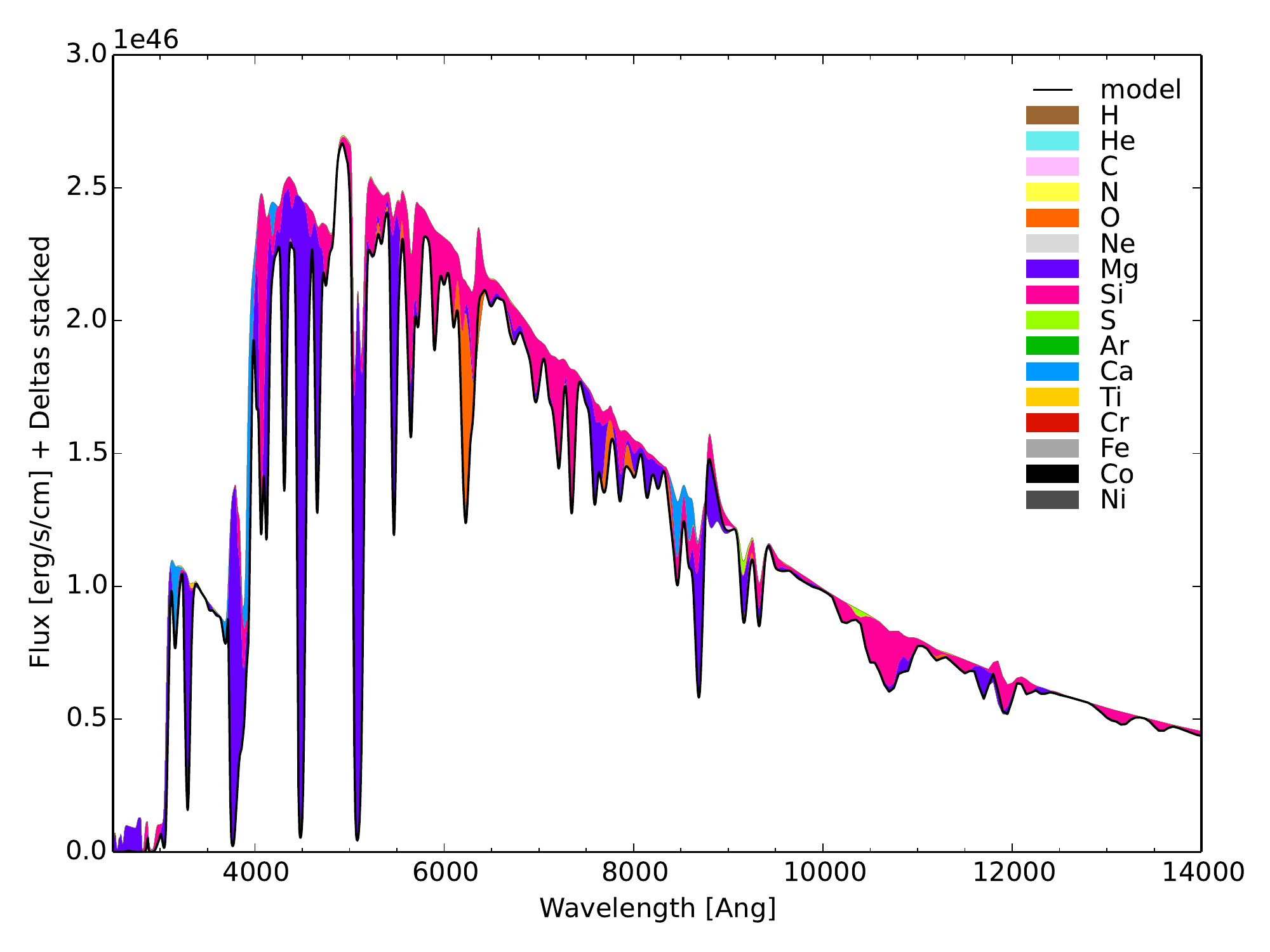}
 \includegraphics[width=.5\textwidth, clip]{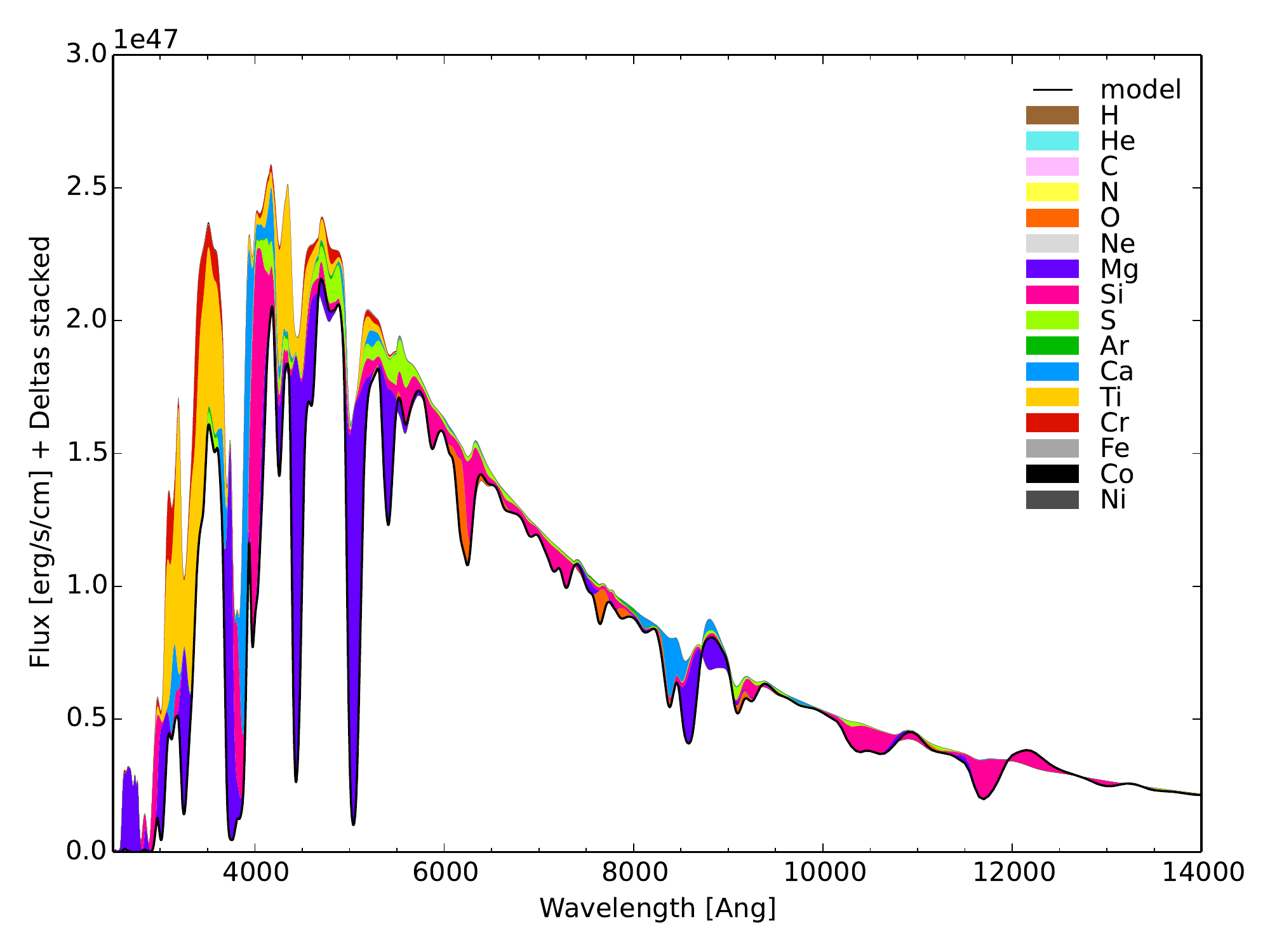}
}
\centerline{
 \includegraphics[width=.5\textwidth, clip]{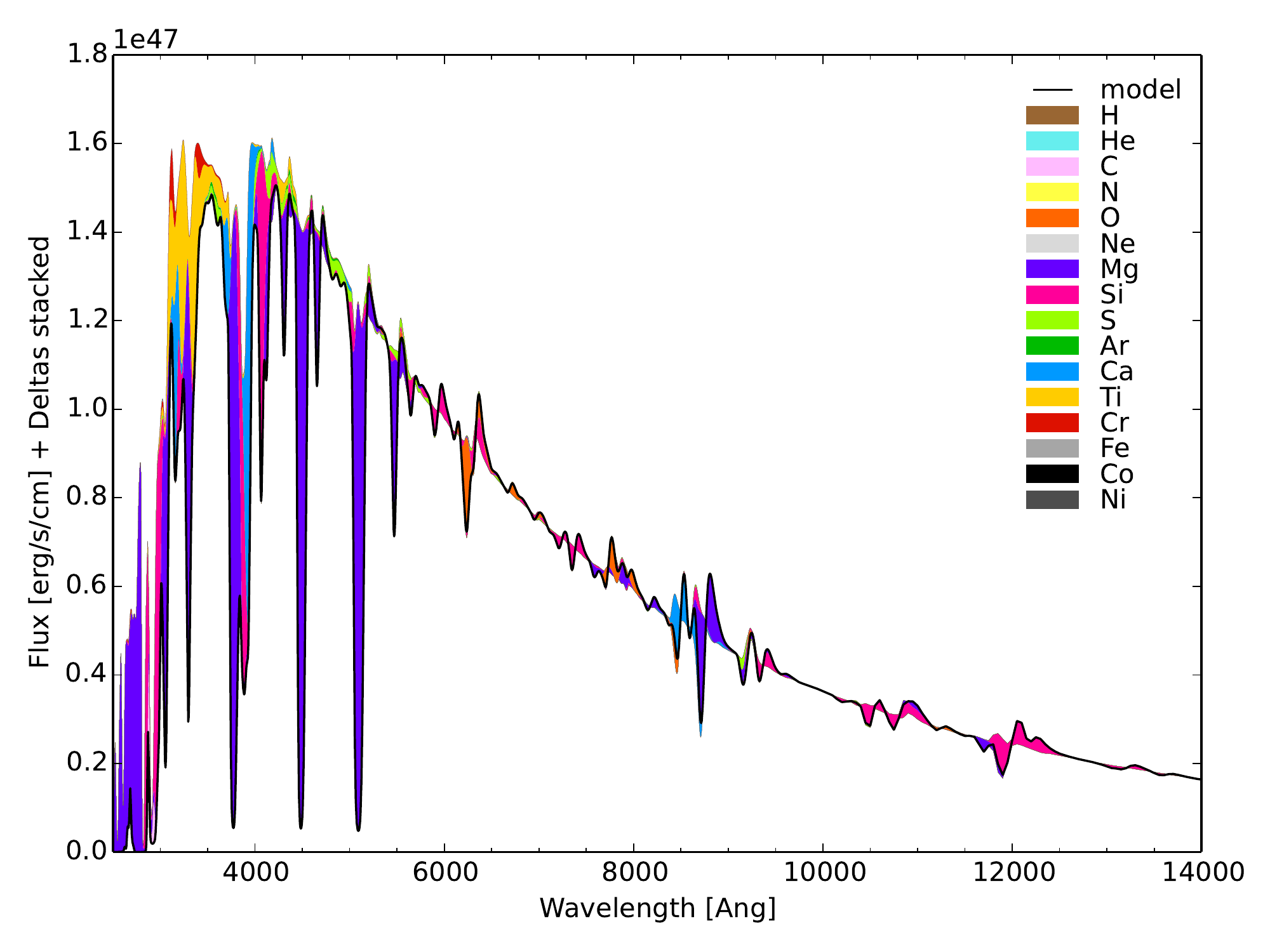}
 \includegraphics[width=.5\textwidth, clip]{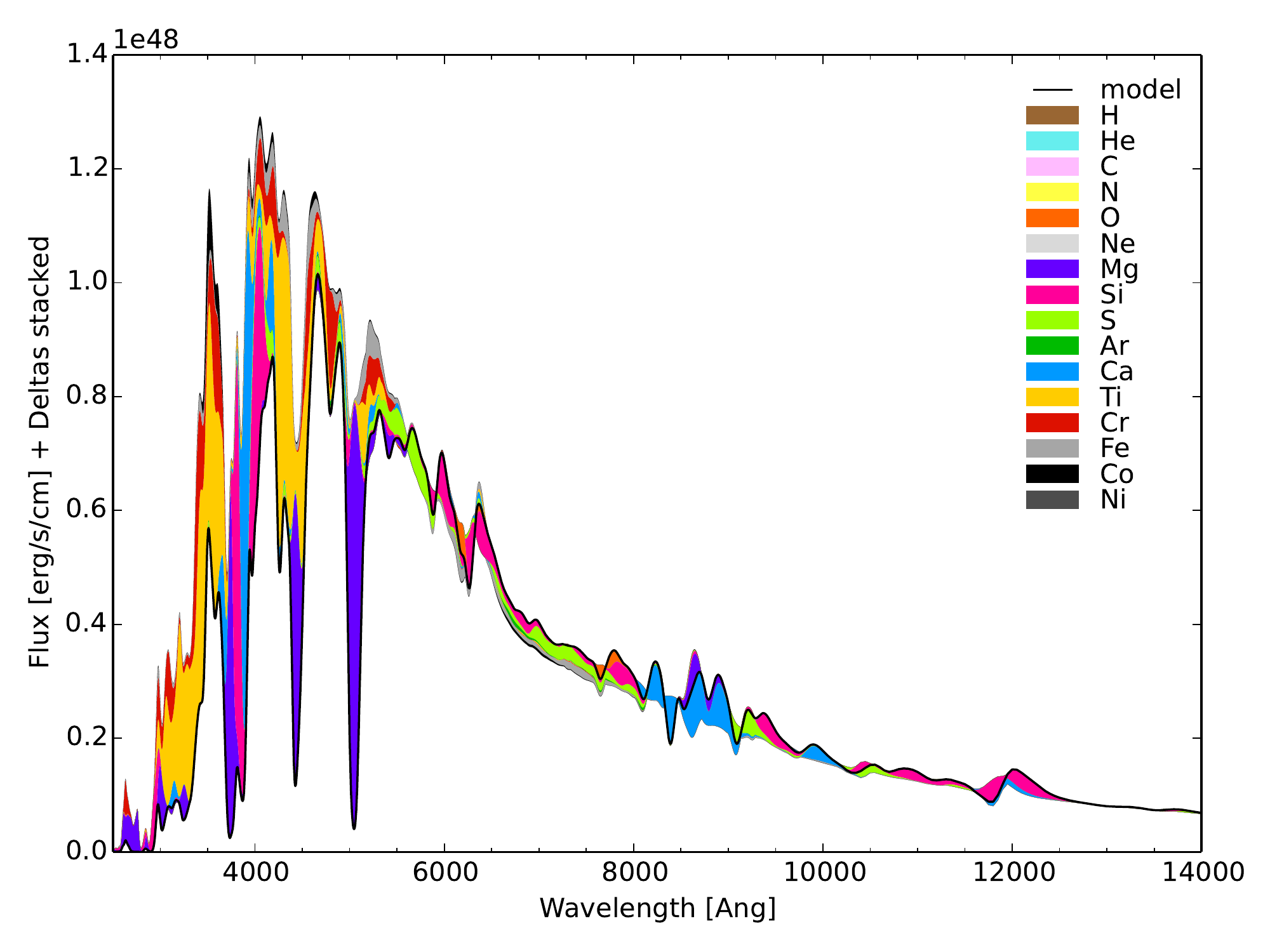}
}
\centerline{
 \includegraphics[width=.5\textwidth, clip]{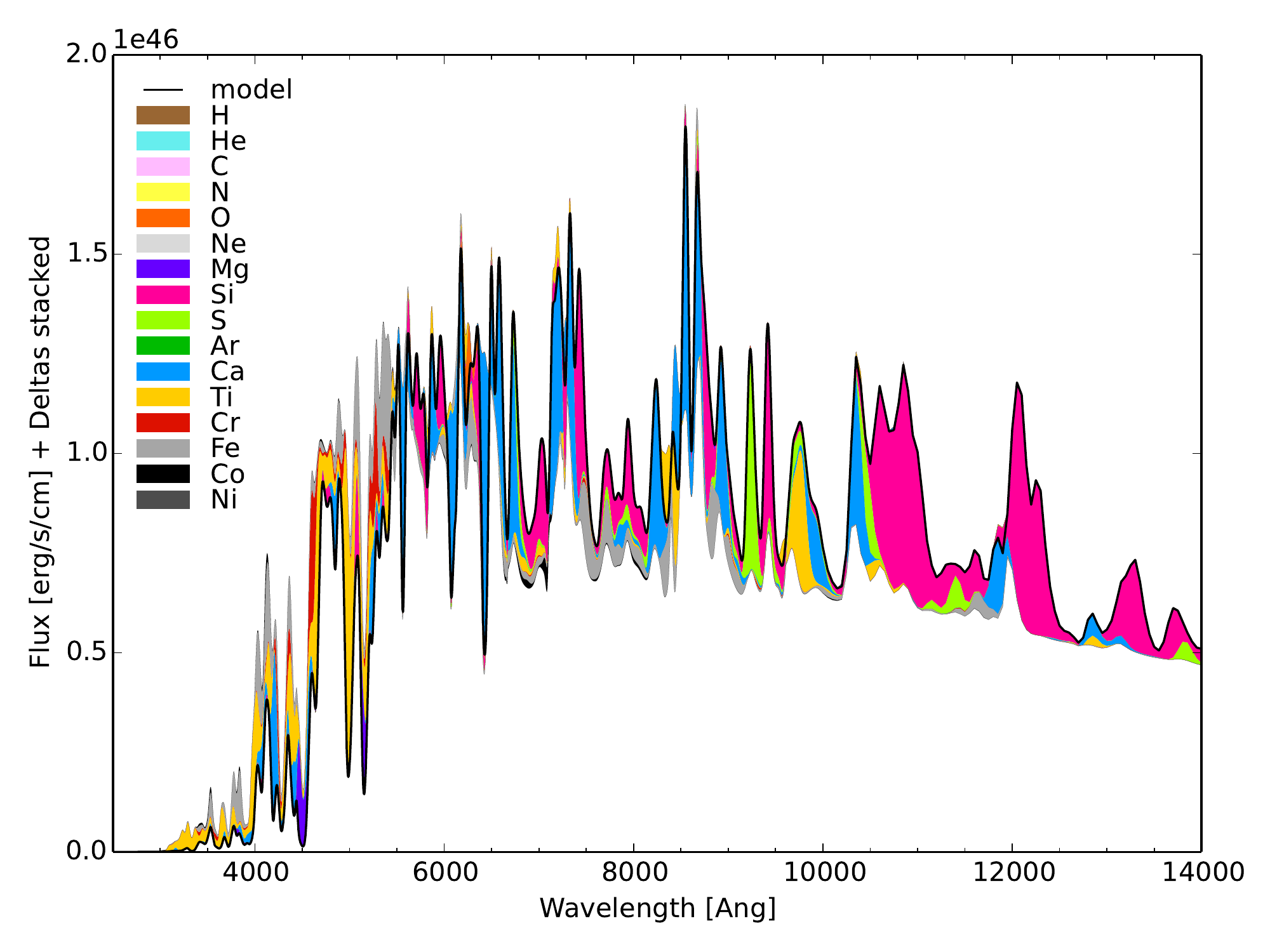}
 \includegraphics[width=.5\textwidth, clip]{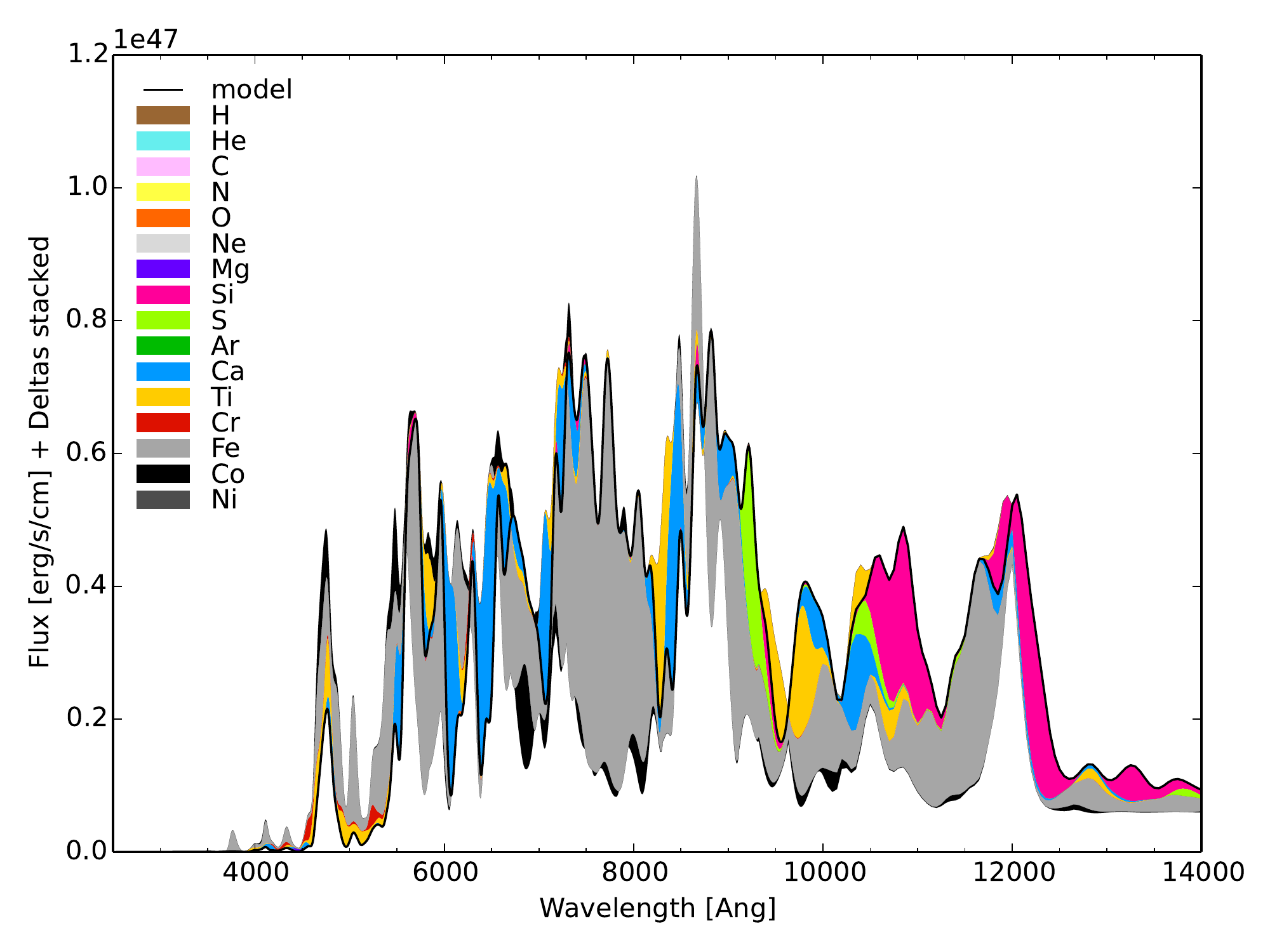}
}
\caption{{\it PHOENIX} spectral line identification for the 90~$M_{\odot}$ model (left panels; -64~d, 0 and 204~d with respect
to peak luminosity from top to bottom) and the
140~$M_{\odot}$ panel (right panels; -84~d, 0 and 184~d with respect to peak luminosity from top to bottom ).
The thick black curve is the original full spectrum. For a more detailed interpretation of this plot
refer to the discussion in \S~\ref{Sec:LTEZ0}.\label{Fig:LTE_LineID}}
\end{figure*}

The  spectral evolution of the lower mass (90~$M_{\odot}$) and the higher mass (140~$M_{\odot}$)
rotating PISN is shown in Figure~\ref{Fig:LTE_specevol}, where the phases cited are with respect
to the time of peak luminosity during the later phase of $^{56}$Ni heating of the ejecta. 
The hotter, more massive models are less line-dominated
than the less massive ones and exhibit a stronger continuum due to their higher temperatures. 
To better illustrate the spectroscopic characteristics of rotating PISNe 
we also show only the peak luminosity spectra for all models in Figure~\ref{Fig:Peak_LTEspec} as well
as detailed spectral line identifications in Figure~\ref{Fig:LTE_LineID} for the 90 and 140~$M_{\odot}$ models. 

The spectral line identification shown in Figure~\ref{Fig:LTE_LineID}
was performed by removing the line-opacity 
from individual ions, re-running the {\it PHOENIX} models and comparing with the original spectrum.
The difference between each of the "knock-out" spectra and the original is attributed to the lines of the respective ion \citep{2012arXiv1208.3781V}. 
In Figure~\ref{Fig:LTE_LineID} we stack the differences in flux level
from different runs on top of each other, using distinct colors for different chemical elements.
Colored patches on top of the original flux level are caused by (the
absence of) absorption, and patches below the original flux level by (the absence of) emission.
The plots show that some of the lines that appear in the spectra can
not be fully attributed to a single chemical element 
but rather to a combined effect of multiple species.

Figures~\ref{Fig:LTE_specevol} through ~\ref{Fig:LTE_LineID} reveal 
that the optical spectra of all rotating PISN spectra are dominated by Intermediate Mass Elements (IMEs). 
More specifically, the early to peak luminosity 
spectra are dominated by strong Mg, Si, and Ca lines with some iron-group elements present
in the near-UV (mostly Cr and Ti). 
The strong Mg absorption is due to the high abundance of Mg in the
middle layer of the PISN ejecta (for example see the upper panel of
Figure~\ref{Fig:140sm_flash}). Enhanced Mg in PISN ejecta
is also seen by other studies \citep{2002ApJ...567..532H, 2007Natur.450..390W} and it is the result of
the passage of the burning flame from this region before it shuts
off prior to reaching the outer ejecta. 
The 6000~\AA~$<\lambda<$~8000~\AA~domain is virtually devoid of prominent line features
in the more massive model (with the exception of the Si $\lambda$6150~\AA).
In addition, all models show prominent Si absorption features in the near-IR ($\lambda >$10500~\AA).

Although the SN ejecta had considerable amounts of He, no significant He features are seen in any
of the spectra due to the fact that the photosphere recedes quickly through the He envelope but also due to the
low envelope temperatures that are insufficient for low atomic level excitation and generation of optical lines.
We caution, however, that non--LTE treatment will likely result in the presence of some He features
due to excitation by secondary electrons produced by Compton scattering of the gamma rays produced
by the radioactive decays of $^{56}$Ni and $^{56}$Co \citep{1991ApJ...383..308L, 2013MNRAS.429.2228H}.
Over time as the expanding SN ejecta
cool and the photosphere recedes into the inner regions additional lines appear characteristic of the
inner ejecta (Ti, Cr, Fe, Co; bottom panels of Figure~\ref{Fig:LTE_LineID}).
The basic spectral characteristics are very similar to those
observed in the He100 models of \citet{2011ApJ...734..102K} and \citet{2013MNRAS.428.3227D}.

\begin{figure}
\begin{center}
\includegraphics[angle=-90,width=9cm,trim=1.in 0.15in 0.5in 0.15in,clip]{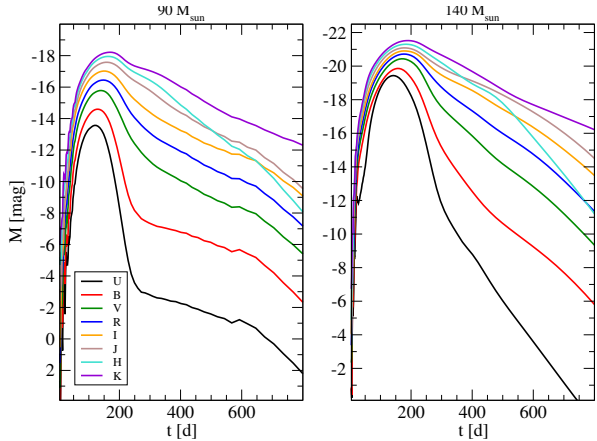}
\caption{Broad-band {\it UBVRIJHK PHOENIX} LCs for the 90~$M_{\odot}$ (left panel) and the 140~$M_{\odot}$ (right panel) models.
Note how the SN LC is dimmer and has a steeper post-maximum decline in shorter wavelengths.
\label{Fig:PhoenixBand_LTELCs}}
\end{center}
\end{figure}

\begin{figure}
\begin{center}
\includegraphics[angle=-90,width=9cm,trim=1.in 0.25in 0.5in 0.15in,clip]{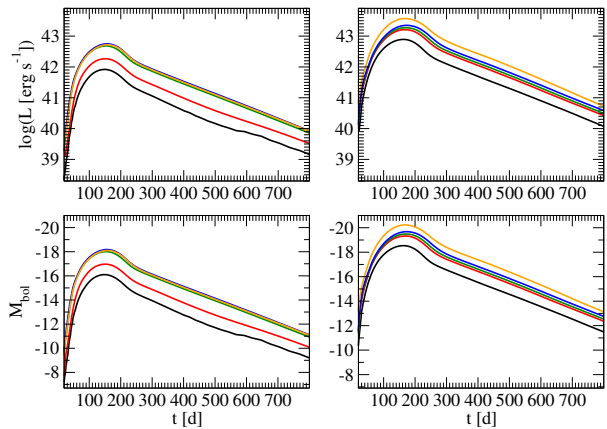}
\caption{{\it PHOENIX} LCs in terms of luminosity (upper panels) and absolute
bolometric magnitude, $M_{\rm bol}$ (lower panels) for the zero metallicity 90-140~$M_{\odot}$ PISNe. 
None of these events is super-luminous and all peak luminosities are within the range of those of
regular core-collapse SNe.
The colors of the curves correspond to the same models as in Figure~\ref{Fig:Full_LCs}.\label{Fig:Phoenix_LTELCs}}
\end{center}
\end{figure}

Convolution of the  spectra with the standard {\it UBVRIJHK} broad-band filters allows for the calculation
of photometric LCs that are shown in Figure~\ref{Fig:PhoenixBand_LTELCs} for the 90~$M_{\odot}$ and the 
140~$M_{\odot}$ models. In a manner characteristic of regular SN events, the redder LCs are brighter
and with longer diffusion time-scales. Integration of the spectra yields the final bolometric
LCs for all of the rotating PISN models studied here, shown in Figure~\ref{Fig:Phoenix_LTELCs}. 
The peak bolometric luminosities are in the range $10^{43}$~-$5 \times 10^{44}$ (corresponding to
bolometric magnitudes in the range -16~$\lesssim M_{\rm bol} \lesssim$~-20) while the rise time
to peak luminosity is in the range 140-170~d in all cases. As such, none of these models are as bright
as typical SLSN events ($M_{\rm bol} >$~-21~mag; \citealt{2012Sci...337..927G}). Rotating PISNe luminosities span
the whole range that is characteristic of other, regular luminosity events such as Type IIP,
Type IIb/c and regular Type IIn events and can also bridge the gap between regular CCSNe
and SLSNe. Rotating PISNe of much larger mass, however, can possess peak luminosities characteristic
of SLSNe (Section~\ref{Sec:LTErotZ}). A distinguishing characteristic of PISN LCs as compared to LCs of other SN events is obviously
the much slower time-scales over which they evolve. 

\begin{figure}
\begin{center}
\includegraphics[angle=-90,width=9cm,trim=0.9in 0.25in 0.5in 0.15in,clip]{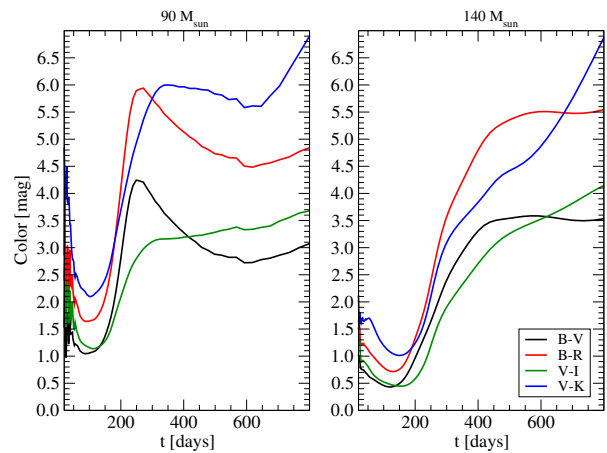}
\caption{Evolution of the $B-V$, $B-R$, $V-I$ and $V-K$ colors for the 90~$M_{\odot}$ (left panel)
and the 140~$M_{\odot}$ (right panel) model. The color evolution of PISNe is redder than
that of other types of SNe.\label{Fig:LTE_ColorEvol}}
\end{center}
\end{figure}

Another characteristic of PISN based upon our analysis is their intrinsically red color regardless of rotation.
Figure~\ref{Fig:LTE_ColorEvol} shows the $B-V$, $B-R$, $V-I$ and $V-K$ color evolution for the
90~$M_{\odot}$ and the 140~$M_{\odot}$ models. All of our models remain red in color, more so during
their post-peak luminosity evolution. As expected, lower mass PISNe are redder than higher mass events.
This is also in agreement with the results of \citet{2013MNRAS.428.3227D}.

\subsection{{\it Effects of rotation and metallicity.}}\label{Sec:LTErotZ}

As discussed in \S~\ref{MESApreSN}, in order to assess the effects of rotation to the radiative properties of PISN, we have run a
rapidly rotating model without the effects of the ST dynamo (140~$M_{\odot}$, no-ST) and 
two non-rotating models (200~$M_{\odot}$ and 260~$M_{\odot}$) all at zero metallicity. The
main properties of these models are presented in Table~\ref{T1}. 

Prior to examining how  LCs and spectra compare between progenitors with different ZAMS rotation rates, we 
summarize the main effects of rotation on the structure of PISN progenitors discussed by \citet{2012ApJ...748...42C}.
Rotationally--induced mixing (mainly due to meridional circulation and the ST mechanism) chemically homogenizes 
the star and enhances production of C, N and O. In addition, rotationally--induced
mass--loss and enhanced radiatively-driven mass--loss due to the increased presence of metals in the outer layers
leads to the loss of all of the H and most (or in some cases all) of the He envelope.
In the case of zero rotation, we
find that PISN progenitors retain both H and He and have RSG-type large radius envelopes. 
The CO core structures
of the stars are, however, very similar regardless of the initial rotation. 
As a result, once the PISN sets in the nucleosynthetic products of the explosion
are similar in all progenitors and the structures imported in {\it PHOENIX} quite similar, regardless of the initial
rotation rate. The only notable differences are the increased surface abundances of C, N and O for the rotating
models. 

\begin{figure}
\begin{center}
\includegraphics[angle=-90,width=9cm,trim=0.9in 0.25in 0.5in 0.15in,clip]{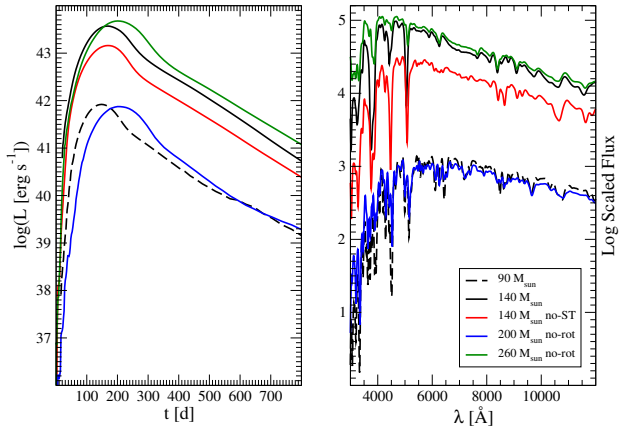}
\caption{Bolometric LCs (left panel) and spectra (right panel) of the rapidly rotating 140~$M_{\odot}$ non magnetic 
(no-ST) model (red curve) and the non-rotating 200 and 260~$M_{\odot}$ models (blue and green curves respectively, \S~\ref{Sec:LTErotZ}) compared to
the original 90~$M_{\odot}$ and 140~$M_{\odot}$ models (solid and dashed black curves respectively).\label{Fig:Rot_effects}}
\end{center}
\end{figure}

For the compact rotating PISN, the photospheric temperatures remain very high over time-scales past
peak luminosity ($>$~10,000~K) keeping H, He, C, N and O ionized. This fact, together
with the intrinsically weak nature of the C and N lines, hints that we should not expect strong lines
indicative of the increased C, N, O surface abundances in the model spectra of the rotating models. 
Figure~\ref{Fig:Rot_effects} showing bolometric LCs (left panel) and peak luminosity spectra (right panel) for models of different
ZAMS rotation rates confirms this assertion.

\begin{figure}
\begin{center}
\includegraphics[angle=-90,width=9cm,trim=0.82in 0.25in 0.5in 0.15in,clip]{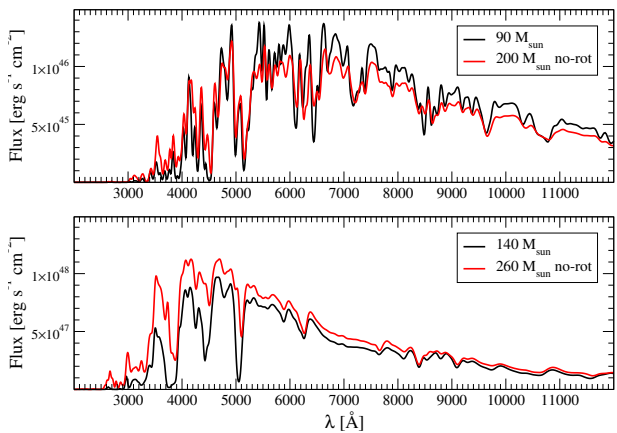}
\caption{Peak spectrum comparisons for the 90~$M_{\odot}$ and the 200~$M_{\odot}$ no-rot (upper panel)
and the 140~$M_{\odot}$ and the 260~$M_{\odot}$ (lower panel) models. The effects of pre--SN rotation
do not have observable effects on the resulting spectra of PISN with similar $M_{\rm CO}$ and bolometric LC properties. 
The enhanced mass--loss suffered by the 90~$M_{\odot}$ model accounts for the more pronounced differences (mainly the \ion{O}{1} 
$\lambda$~6454~\AA~and $\lambda$~7777~\AA absorption features).\label{Fig:peak_rot_spec}}
\end{center}
\end{figure}

We observe that the bolometric LCs of the 90~$M_{\odot}$ and the 200~$M_{\odot}$ models and those of the
140~$M_{\odot}$ and the 260~$M_{\odot}$ models have similar peak luminosities (albeit different diffusion time-scales)
and we will therefore focus our specroscopic comparison efforts to these two groups of models. A more clear comparison
of the peak spectra for these two sets of models is shown in the lower panel of Figure~\ref{Fig:peak_rot_spec}.
The main spectral features of the 140~$M_{\odot}$ and 260~$M_{\odot}$ models are very similar considering the different peak
bolometric luminosities.
Strong Si, Mg and Ca features are seen in the optical with Mg and iron-group line blends strongly present in the near-UV. 
Regardless of the intrinsic differences in the temperature and the velocity of the expanding SN photospheres (leading to differences in
the line fluxes in the 3000-5000~\AA region), there are no
clear indicators of pre--SN rotation present in these spectra. The spectra at earlier phases also appear to be very similar
between the two models.

\begin{figure*}
\begin{center}
\includegraphics[angle=-90,width=16cm,trim=0.82in 0.25in 0.5in 0.15in,clip]{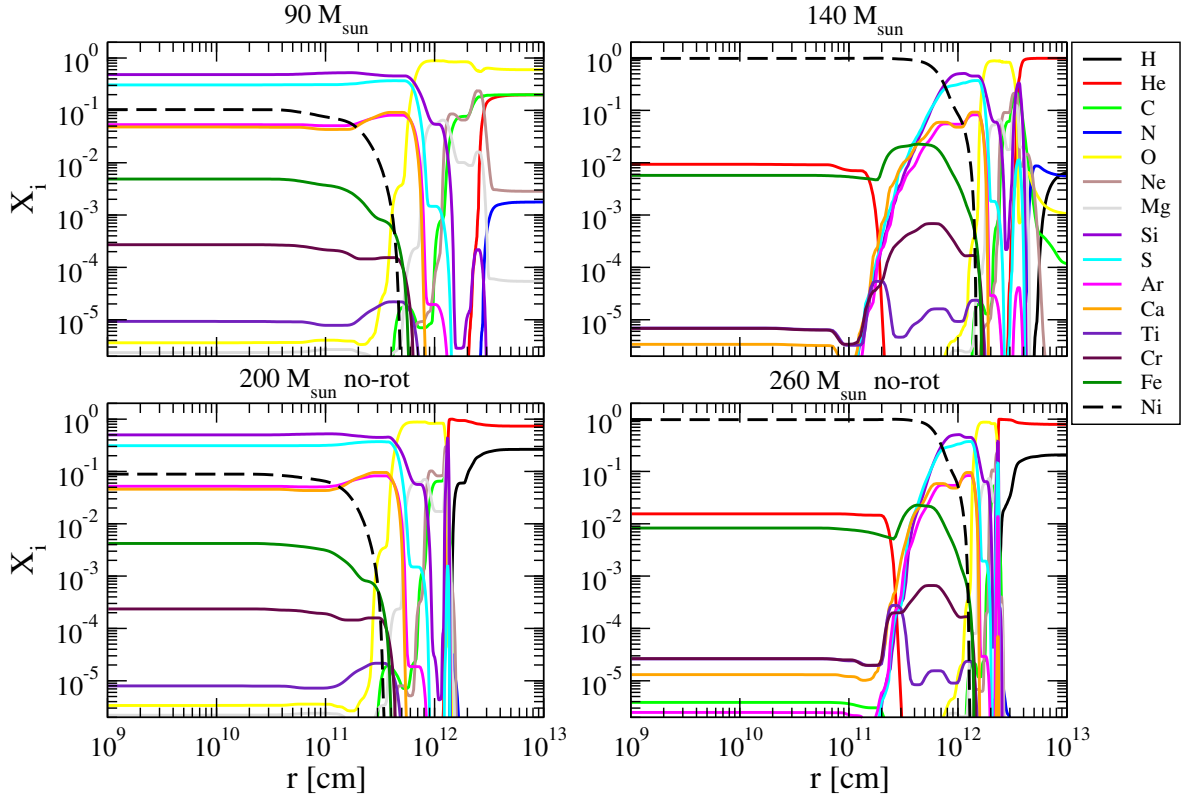}
\caption{Comparison of the pre-SN compositions of the rotating 90~$M_{\odot}$ and 140~$M_{\odot}$ models (upper panels)
against the non-rotating 200~$M_{\odot}$ and 260~$M_{\odot}$ models (lower panels) that produce similar bolometric
LCs (Fig~\ref{Fig:Rot_effects}).\label{Fig:rot_comp_comp}}
\end{center}
\end{figure*}

In the 90~$M_{\odot}$ versus 200~$M_{\odot}$ case there are a few notable differences attributable to the pure CO core
structure of the 90~$M_{\odot}$ PISN. Mass--loss led to the total loss of the H and He envelope for this model leaving
an oxygen-rich progenitor star prior to explosion. In contrast, the outer layers of the 200~$M_{\odot}$ model were still
H and He rich. As a result the peak luminosity spectrum of the 90~$M_{\odot}$ model shows stronger \ion{O}{1} absorption
features at $\lambda$~6454~\AA~and $\lambda$~7777~\AA. These subtle differences can be more clearly distinguished in
the upper panel of Figure~\ref{Fig:peak_rot_spec}. 
For better intuition we show the SN ejecta compositions of the 
rotating and non-rotating models compared in Figure~\ref{Fig:rot_comp_comp}.
In can be seen that the more massive models (right panels; 140~$M_{\odot}$ and 260~$M_{\odot}$) have similar SN ejecta 
composition profiles regardless of rotation while the lower mass models have differences in the outer ejecta with the 90~$M_{\odot}$
model having an O-rich outer envelope.

\begin{figure}
\begin{center}
\includegraphics[angle=-90,width=9cm,trim=0.9in 0.25in 0.5in 0.15in,clip]{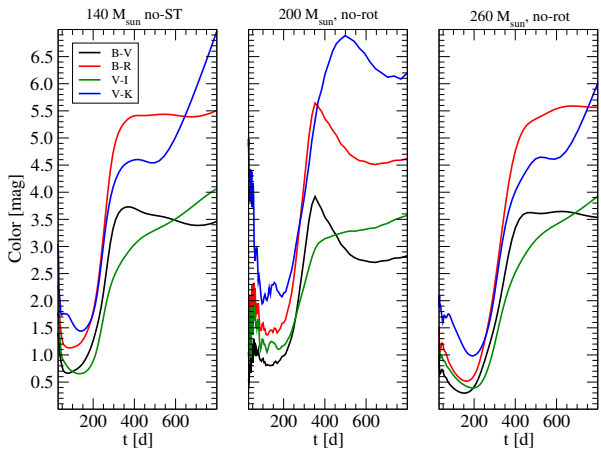}
\caption{Evolution of the $B-V$, $B-R$, $V-I$ and $V-K$ colors for the 140~$M_{\odot}$ non-magnetic (no-ST; left panel),
the 200~$M_{\odot}$ no-rot (middle panel) and the 260~$M_{\odot}$ no-rot (right panel) model (\S~\ref{Sec:LTErotZ}).\label{Fig:Color_rot}}
\end{center}
\end{figure}

The color evolution for the models of different ZAMS rotation rates
discussed here is shown in Figure~\ref{Fig:Color_rot}. A comparison of this figure with Figure~\ref{Fig:LTE_ColorEvol} indicates that,
based on the color properties alone, one cannot see a clear distinction between otherwise similar
rotating and non-rotating PISN.

To infer the effects of rotation on the radiative properties of PISN, and more specifically on spectra, one has to be
cautious as to not over-interpret the observations. The features in the optical part of the spectrum are more robust
than the ones in the near-UV where Mg but also a variety of iron-group elements (Ti, Cr, Fe) contribute to several
line-blends. Even in the case of the \ion{O}{1} features present in the model spectrum of the 90~$M_{\odot}$ PISN, it
would be unfounded to attribute them solely to the effects of rotation since mass--loss was the primary reason the progenitor
star was mostly composed of pure oxygen. Although rotation is one avenue of mass--loss enhancement there are
alternative possibilities (binary interaction, episodic/Luminous Blue Variable (LBV-type) mass--loss). Also, differences
in the SN blast profiles (in terms of temperature and velocity) further increase the degeneracy between non-rotating
and rotating PISN spectra. Perhaps the only qualitative difference between rotating and non-rotating models is the
apparent suppression of the near-UV flux ($\lambda <$~3500~\AA) attributable to the increased presence of metals
in the rotating case due to the effects of pre--SN mixing. Conversely, this effect could also be due to intrisically
higher metallicity, regardless of rotation \citep{2011ApJ...734..102K}. 

\begin{figure}
\begin{center}
\includegraphics[angle=-90,width=9cm,trim=0.9in 0.25in 0.5in 0.15in,clip]{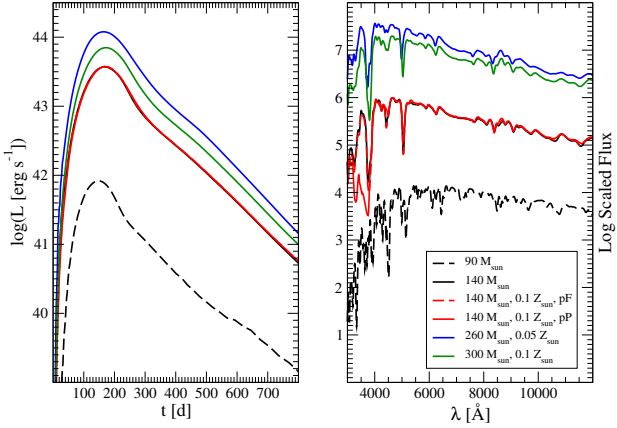}
\caption{Bolometric LCs (left panel) and spectra (right panel) of the 0.1~$Z_{\odot}$ 140~$M_{\odot}$ model
prior to import to {\it FLASH} (``pF''; dashed red curve) and {\it PHOENIX} (``pP''; solid red curve),
the 0.05~$Z_{\odot}$~260~$M_{\odot}$ and the 0.1~$Z_{\odot}$~300~$M_{\odot}$ model (blue and green curves respectively) compared to
the original 90~$M_{\odot}$ and 140~$M_{\odot}$ models (solid and dashed black curves respectively).
Evolution with different initial metallicity does not change the resulting PISN spectra significantly. Artifical addition
of metal content prior to the radiative transfer calculation leads to the suppression of UV flux (model ``pP'').\label{Fig:Z_effects}}
\end{center}
\end{figure}

\begin{figure}
\begin{center}
\includegraphics[angle=-90,width=9cm,trim=0.9in 0.25in 0.5in 0.15in,clip]{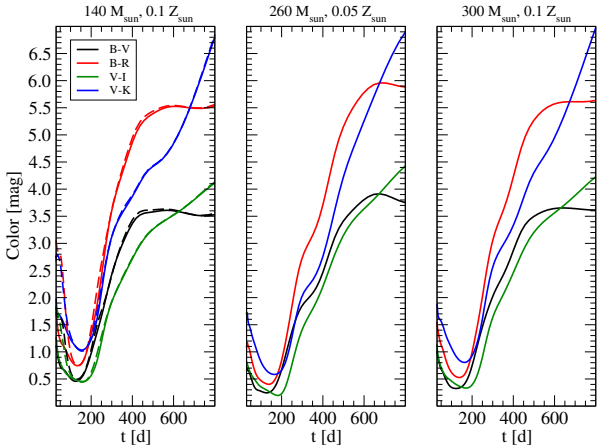}
\caption{Evolution of the $B-V$, $B-R$, $V-I$ and $V-K$ colors for the 0.1~$Z_{\odot}$~140~$M_{\odot}$ (left panel),
the 0.05~$Z_{\odot}$~200~$M_{\odot}$ (middle panel) and the 0.1~$Z_{\odot}$~300~$M_{\odot}$ (right panel) 
model (see Section~\ref{Sec:LTErotZ}). The dashed curves in the left panel correspond to the 140~$M_{\odot}$ ``pP'' model.\label{Fig:Color_z}}
\end{center}
\end{figure}

To study the effects of metallicity on the radiative properties of PISNe we have run the models
0.05~$Z_{\odot}$~260~$M_{\odot}$, 0.1~$Z_{\odot}$~300~$M_{\odot}$ and 0.1~$Z_{\odot}$~140~$M_{\odot}$
as described in \S~\ref{MESApreSN} and Table~\ref{T1}.
Figure~\ref{Fig:Z_effects} shows the resulting {\it PHOENIX} bolometric LCs (left panel) and spectra (right panel) as compared
to our base zero metallicity 90~$M_{\odot}$ and 140~$M_{\odot}$ models. Figure~\ref{Fig:Color_z} shows the
color evolution for these higher metallicity models. 

\begin{figure}
\begin{center}
\includegraphics[angle=-90,width=9cm,trim=0.9in 0.25in 0.5in 0.15in,clip]{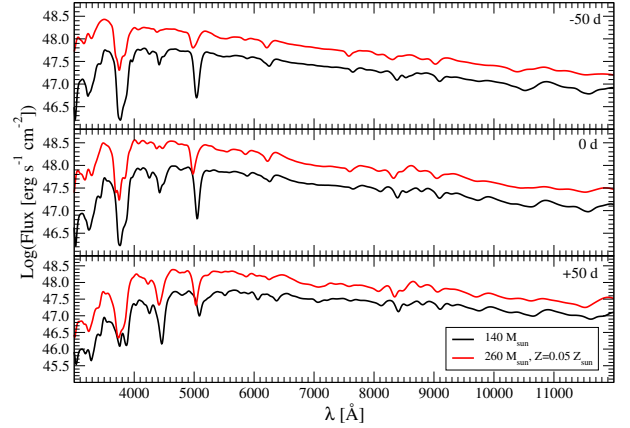}
\caption{Comparison of the 140~$M_{\odot}$ and the 0.05~$Z_{\odot}$~260~$M_{\odot}$ model spectra
at the contemporaneous epochs of -50~d (upper panel), 0~d (middle panel) and +50~d (lower panel)
with respect to peak luminosity. There are no significant differences induced by different ZAMS
metallicity in the spectral evolution of otherwise similar PISNe.\label{Fig:140vs260z005}}
\end{center}
\end{figure}

The peak luminosity spectra for rotating PISN evolved with different metallicities at ZAMS (solid black, blue and green curves
in the right pane of Figure~\ref{Fig:Z_effects}) all appear to be very similar with no distinguishable differences in any
particular spectral features. This seems to be the case independently of phase as we illustrate in Figure~\ref{Fig:140vs260z005},
where the rotating zero metallicity 140~$M_{\odot}$ model spectra are compared with those of the
0.05~$Z_{\odot}$~260~$M_{\odot}$ model at the same phases (before, during and after peak luminosity).

In contrast, for the 140~$M_{\odot}$ models for which a metal content corresponding to 0.1~$Z_{\odot}$ was
added prior to import to {\it FLASH} (``pF'' model; dashed red curve) and {\it PHOENIX} (``pP'' model; solid red curve) differences are seen in the near-UV
spectral flux and Ca H\&K absorption. For details on these models refer back to the discussion in \S~\ref{MESApreSN}.
More specifically, the near-UV flux is suppressed in the ``pP'' model as compared to the original 140~$M_{\odot}$ model.
The ``pF'' model also shows a much smaller far UV flux suppression at $\lambda <$~4500~\AA~while for the most part
the spectrum is identical to the original 140~$M_{\odot}$ model.
The flux deficit is more dramatic in the ``pP'' case where the metallicity was added prior to calculating the {\it PHOENIX}  spectra. 
This is because once the PISN explosion occur, the bulk of the additional metal content in the core of the progenitor contributes
to the formation of $^{56}$Ni and other iron-group elements through nuclear burning.
By the time the SN photosphere expands, the blast profile is very similar to that of the 140~$M_{\odot}$ model. 
In other words, the progenitor metallicity information is undecipherable in the post-explosion PISN spectra.

This is an important point of contrast with the results of \citep{2011ApJ...734..102K} illustrated in their Figure 11. They too find
that increased metallicity has the effect to reduce the UV flux for otherwise identical PISN progenitors, but they also add the metal
content immediately prior to doing the {\it SEDONA} radiative transfer calculations. Self-consistent evolution of PISNe with higher
metal content from the ZAMS or even addition of extra metal abundances to a model prior to the explosion does not lead to significant 
differences in the spectra of our rotating models as compared to zero or low-metallicity models. Note that, 
as we discuss in Section~\ref{Sec:Comp_other}, our PISN models compare well with the
\citep{2011ApJ...734..102K} models of the same ZAMS metallicity.

The radiative properties of PISNe appear to be degenerate across differences in ZAMS rotation rate and metallicity 
and seem to depend more on basic
SN blast profile properties such as the temperature and velocity of the SN ejecta. Once the explosion 
sets in, the products of nuclear burning are predominantly IMEs in the outer and iron-peak
elements in the inner ejecta. Even the detailed spectral evolution of PISNe appears to be quite similar for models
of different initial conditions. Pre-SN surface composition, on the other hand, seems to be the only factor leading
to robust differences in the PISN spectra, as showcased by the 90~$M_{\odot}$ model discussed above.
Differences in pre--SN composition are primarily attributable to the extent of mass--loss suffered by the progenitor.
As such, H/He-poor PISNe can be distinguished by the presence of \ion{O}{1} absorption features in the optical,
and that only if the appropriate temperature conditions are maintained in the SN ejecta. 

\subsection{{\it Comparison with other PISN studies.}}\label{Sec:Comp_other}

The  LCs and color evolution we find for rotating PISNe have similar broad, qualitative properties
with the non--LTE LCs of the non-rotating models studied by \citet{2011ApJ...734..102K} and \citet{2013MNRAS.428.3227D}.
To compare with these studies, we use our brightest zero metallicity PISN model (140~$M_{\odot}$). This model
is the closest to the bare ``helium'' star He100 models of both \citet{2011ApJ...734..102K} (He100K) and \citet{2013MNRAS.428.3227D} (He100D) 
in terms of $M_{\rm CO}$, $M_{\rm Ni}$ and final structural properties modulo differences in the surface abundances of C, N and O that are enhanced
in our model due to the effects of rotationally-induced mixing.

The bolometric LC of our 140~$M_{\odot}$ model
is very close in terms of both peak luminosity ($M_{\rm bol,peak} =$~-20.2~mag, $L_{peak} \simeq 4 \times 10^{43}$~erg)
and rise time to maximum ($t_{peak} \simeq$~170~d) with the He100K and He100D models.
The same holds for the broad band LCs: the same behavior is seen in all three studies going 
from the bluer ({\it U}) to the redder ({\it K}) band. An exception is the range of peak filter
magnitudes where our model is more in agreement with He100D than with He100K 
who find a larger
range ($\sim$~-19~$<M<$~-26 from {\it U$_{\rm peak}$} to {\it K$_{\rm peak}$}). The peak absolute
{\it V} magnitudes recovered from all three studies are in good agreement ($M_{V} \sim$~-20.5 to -21)
regarding the differences in the initial progenitors. 
The  color evolution of the 140~$M_{\odot}$ model is qualitatively 
similar to that of the He100D model, although we find it to be consistently
redder for the rotating PISN. This could be due to differences between our LTE treatment
and the \citet{2013MNRAS.428.3227D} non--LTE approach.

\begin{figure}
\begin{center}
\includegraphics[angle=-90,width=9cm,trim=0.9in 0.25in 0.5in 0.15in,clip]{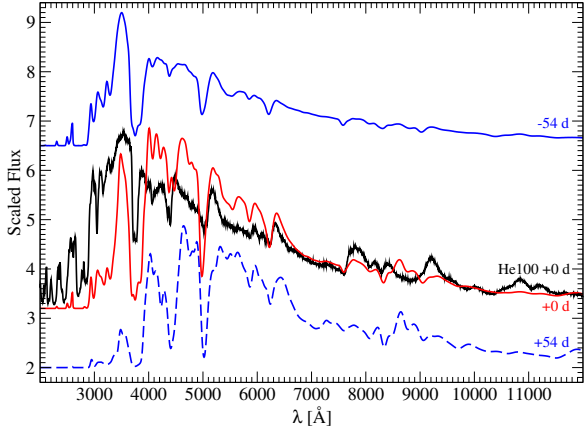}
\caption{Comparison of the \citet{2011ApJ...734..102K} He100 peak luminosity spectrum (black curve)
with the -54~d, +54~d (solid and dashed blue curve respectively) and peak luminosity (red curve) spectra
of the 140~$M_{\odot}$ model.\label{Fig:Kasen_comp}}
\end{center}
\end{figure}

A comparison of the spectral evolution between the three studies reveals several similarities
as well. In terms of the peak luminosity spectrum, the strongest features seem
to be recovered by all three approaches, somewhat resembling Type Ic SN spectra. 
More specifically the strong \ion{Mg}{1}, \ion{Mg}{2}, \ion{Si}{2}, \ion{Ca}{2} and \ion{O}{1} lines
are recovered by all studies. Figure~\ref{Fig:Kasen_comp} shows a comparison of
the 140~$M_{\odot}$ spectra for different phases with respect to peak luminosity
and the He100K peak spectrum. 

Given the intrinsic differences of the progenitors, the agreement between the spectra of the 140~$M_{\odot}$ model
and He100K is very good in the optical (Figure~\ref{Fig:Kasen_comp}). Obvious differences do exist in some near-IR features ($\lambda >$~8300~\AA)
and the near-UV flux. We wish to note that besides examining the radiative properties of PISN this excercise also serves as
direct comparison of model PISN spectra obtained from different codes. It is remarkable that given differences in the
numerical treatment of radiative transfer between {\it PHOENIX} and {\it SEDONA} (that uses a Monte Carlo method) the
main PISN spectroscopic characteristics recovered are in good agreement. 

\subsection{Comparison with observations.}\label{Comps}

\begin{figure}
\begin{center}
\includegraphics[angle=-90,width=9cm,trim=1.in 0.25in 0.5in 0.15in,clip]{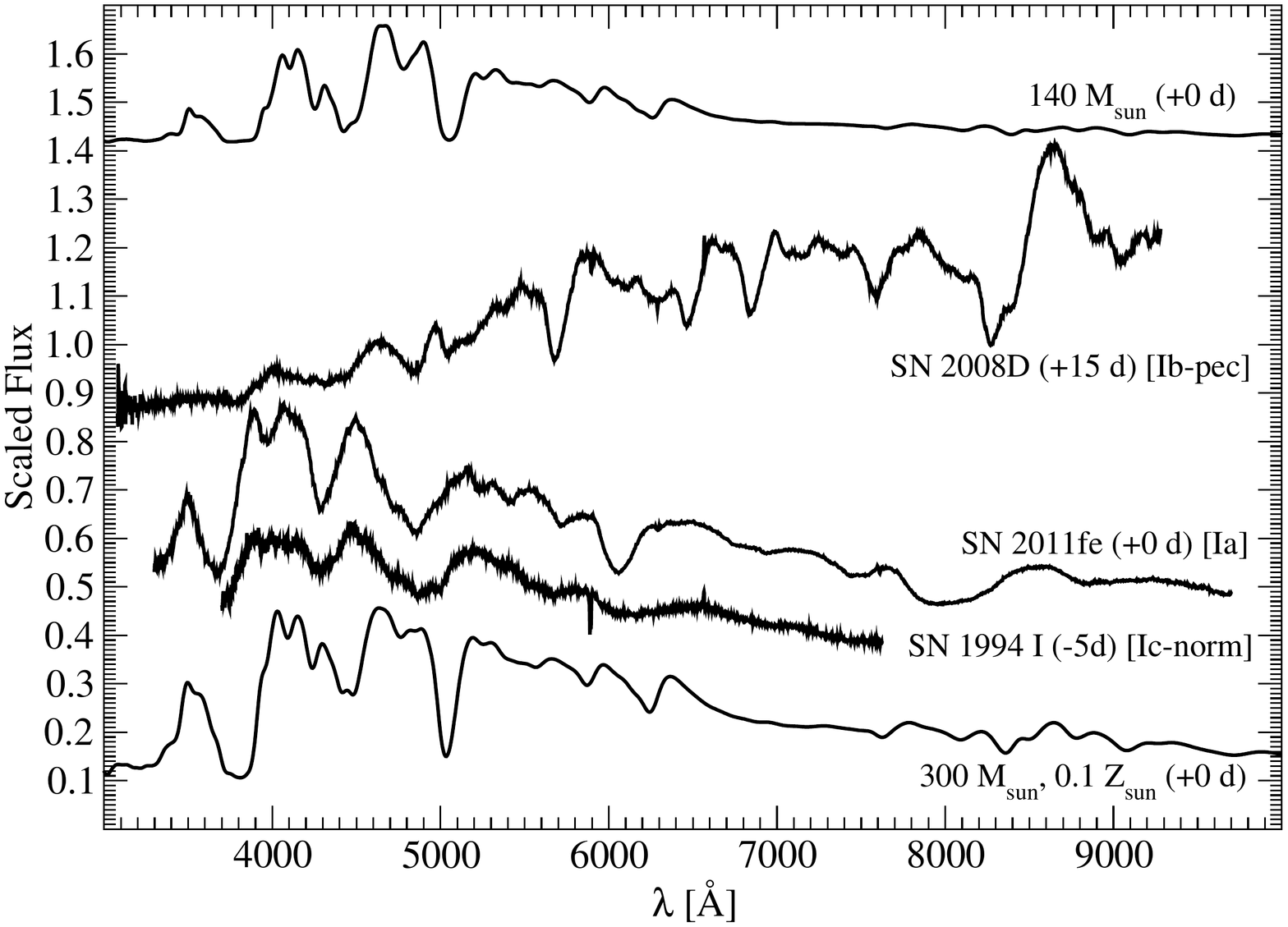}
\caption{Comparison of the 140~$M_{\odot}$ and the 300~$M_{\odot}$ ($Z =$~0.1~$Z_{\odot}$) 
rotating PISN model spectra  at peak luminosity with 
observed spectra of the the normal Type Ic
SN~1994I \citep{1995ApJ...450L..11F}, the Type Ia SN~2011fe \citep{2011Natur.480..344N} and the peculiar
Type Ib SN~2008D \citep{2009ApJ...702..226M}. The parentheses indicate the phase after
peak luminosity in days and the brackets the SN type. All the observed 
SN spectra were downloaded from the Weizmann Interactive
Supernova Data Repository ({\it WISeREP}; \citealt{2012PASP..124..668Y}).\label{Fig:LTE_datacomp}}
\end{center}
\end{figure}

Given the main properties of model PISN spectra that we have discussed in this work, it remains to examine 
if such events have been actually observed. At first glance, the general appearance
of model PISN spectra appears to qualitatively resemble Type Ia and some normal Type Ic SN events with the lack of apparent H and He
features and a strong presence of metals such as O, Mg, Si and Ca. We attempt to illustrate these similarities in
Figure~\ref{Fig:LTE_datacomp} where we plot the peak luminosity 140~$M_{\odot}$ and 0.1~$Z_{\odot}$~300~$M_{\odot}$ spectra against
near-maximum spectra of SN~2008D (Type Ib), SN~2011fe (Type Ia) and SN~1994I (Type Ic). Obviously, SN~2008D is a bad match
(also given the absence of He in our model spectra)
but the other two events exhibit some similarities with model 0.1~$Z_{\odot}$~300~$M_{\odot}$.

\begin{figure}
\begin{center}
\includegraphics[angle=-90,width=9cm,trim=0.9in 0.25in 0.5in 0.15in,clip]{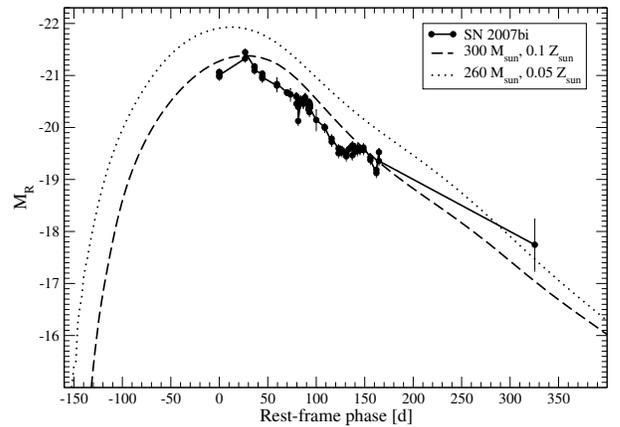}
\caption{The observed R-band LC of SN~2007bi \citep{2009Natur.462..624G} (solid curve) compared
to the R-band model LCs of of the rotating 0.1~$Z_{\odot}$~300~$M_{\odot}$ (dashed curve) and 
0.05~$Z_{\odot}$~260~$M_{\odot}$ (dotted curve) models. Although PISN model LCs seem to be a good
match to the observed LCs of some SLSNe, such as SN~2007bi, the predicted spectral evolution is
remarkably different than the one observed.\label{Fig:07bi_LC_comp}}
\end{center}
\end{figure}

With regards to the proposed association between PISNe and a few observed SLSN events \citep{2009Natur.462..624G},
we find that the emission properties of PISNe are quite different from those of the proposed
candidate SN~2007bi, in agreement with the results of \citet{2013MNRAS.428.3227D}. In particular, we find
that PISNe of several metallicities and rotation rates are intrinsically red events and, in most cases, do not produce
super-luminous events due to the severe mass--loss suffered during the progenitor evolution. This is due to the
fact that PISNe tend to have lower $M_{\rm Ni}/M_{\rm f}$ ratios than Type Ia or Type Ic-norm events leading to
slowly evolving LCs with peak luminosities that span a large range from sub-luminous to some super-luminous
events in the most extreme cases. For very high ZAMS mass ($>$~200~$M_{\odot}$), several solar masses of $^{56}$Ni
are produced powering a super-luminous explosion. From the suite of models considered in our work with metallicities
in agreement with that of the host of SN~2007bi, the only
two that reach peak luminosities comparable to those of the event are the 0.1~$Z_{\odot}$~300~$M_{\odot}$ and the
0.05~$Z_{\odot}$~260~$M_{\odot}$ models. 
The observed R-band LC of SN~2007bi is compared against the LCs of these two models in Figure~\ref{Fig:07bi_LC_comp}.

Based only on the LC, it can be deduced that a rotating PISN with $Z \simeq$~0.05-0.1~$Z_{\odot}$ is indeed
a good model for SN~2007bi; however, a closer look at the spectral and color evolution of this event as
compared with the model predictions reveals certain inconsistencies. 
The most important disagreement stems from a careful comparison of the model PISN spectra with those
of SN~2007bi at contemporaneous epochs. This is an issue originally raised by \citet{2013MNRAS.428.3227D}
with regards to the apparent spectral agreement found between model He100K and the observed
+54~d post-maximum spectrum of SN~2007bi by \citet{2011ApJ...734..102K}. The agreement was due to
the fact that the He100K spectrum compared with the data was one at 50~d before peak, at an epoch
when the PISN was still hot and blue. At later, post-maximum epochs, when the spectra of SN~2007bi were obtained, 
the model PISN spectra are much redder and strongly line-blanketed. 
A caveat to performing nearly-contemporaneous model to data spectral comparisons is the fact that
for many SLSN events, including SN~2007bi, the exact explosion date and therefore the phase
of the observed spectrum are quite uncertain, as noted by \citet{2013ApJ...773...76C}. 
Nevertheless, to be consistent with the other authors, we will also accept the phase of the 
earliest observed SN~2007bi spectrum to be +54~d after peak luminosity.

\begin{figure}
\begin{center}
\includegraphics[angle=-90,width=9cm,trim=0.9in 0.25in 0.5in 0.15in,clip]{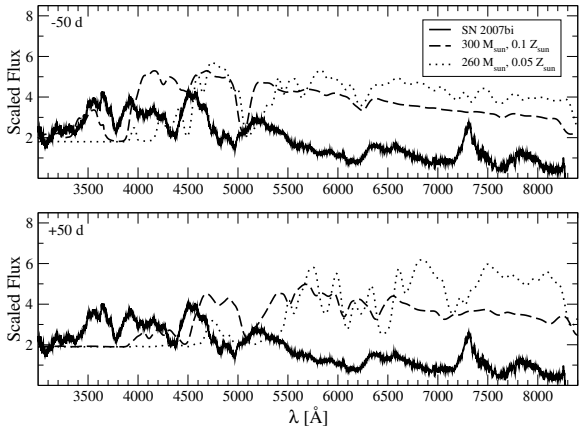}
\caption{Comparison of the spectrum of SN~2007bi at +54d \citep{2009Natur.462..624G} (solid curve) with
that of the 0.1~$Z_{\odot}$~300~$M_{\odot}$ (dashed curve) and the
0.05~$Z_{\odot}$~260~$M_{\odot}$ (dotted curve) model at -50~d (upper panel) and +50~d (lower
panel) with respect to peak luminosity.\label{Fig:07bi_spec_comp}}
\end{center}
\end{figure}

Figure~\ref{Fig:07bi_spec_comp} shows the -50~d and +50~d (with regards to peak
luminosity) spectra of the rotating 0.05~$Z_{\odot}$~260~$M_{\odot}$ and 0.1~$Z_{\odot}$~300~$M_{\odot}$ 
models compared with the observed spectrum of SN~2007bi +54~d after peak luminosity.

The comparison at the nearly contemporaneous epoch of +50~d (lower panel of Figure~\ref{Fig:07bi_spec_comp})
reveals that the PISN models possess continua that are much redder and with different spectral features than the
observations of SN~2007bi suggest. The agreement between the earlier model spectrum of the 0.1~$Z_{\odot}$~300~$M_{\odot}$ model 
(-50~d) and the data is slightly better but yet not nearly strong enough to suggest a physically consistent connection to 
the PISN mechanism. The color evolution of the model PISN spectra further affirms the fact that SN~2007bi was not
a PISN. SN~2007bi retained blue color and a hot continuum for a long time after peak suggesting that a different, more efficient 
SN ejecta heating mechanism was at play and for a longer time-scale than the radioactive decay of $^{56}$Ni. Potential
alternative candidates are magneto-rotational energy injection due to the spin-down of a young magnetar \citep{2012MNRAS.426L..76D} and
H-poor SN ejecta - CSM interaction \citep{2012ApJ...760..154C,2013ApJ...773...76C}. More detailed radiation hydrodynamics and non--LTE radiative transfer
modelling is required to clearly identify the power-input mechanism for SN~2007bi and other H-poor SLSN (SLSN-I).

\section{DISCUSSION AND CONCLUSIONS}\label{Disc}

In this paper we explored the radiative emission properties of model PISNe
for a variety of progenitors of different ZAMS mass (90-300~$M_{\odot}$), 
metallicity (0-0.1~$Z_{\odot}$) and rotation rate.
The analysis we performed was done in three basic stages: stellar evolution from ZAMS
up to pre--SN with {\it MESA}, nuclear burning and 
hydrodynamic evolution of the explosion phase up to post SN shock break out with {\it FLASH}
and radiative transfer with {\it PHOENIX} yielding the final model spectra and LCs. 
For a subset of our models (the zero metallicity rotating 90-140~$M_{\odot}$ models) an
additional step included the mapping of the pre-SBO profiles to the radiation hydrodynamics
code {\it RAGE} where the phase of SN shock breakout and the corresponding LCs and
SEDs were computed. 

As has been discussed in past studies \citep{2012ApJ...748...42C,2012A&A...542A.113Y}
the main effects of rotation on the evolution of PISN progenitors is the enhanced mixing
leading to the formation of larger CO cores for initially smaller ZAMS mass as compared
to the case of zero rotation, and the increased mass--loss suffered leading to
bare CO stars with little or no He in their envelopes prior to explosion. As such, the pre-explosion
zero metallicity models we consider here were compact ($R_{\rm f} < 2 \times 10^{11}$~cm) 
and dense all the way to the surface. The {\it RAGE} simulations revealed that
the SBO SEDs of these events are significantly hotter than those implied
for RSG and BSG progenitors \citep{2011ApJ...734..102K} peaking in the hard X-rays
(1-100~keV). Nevertheless, the combination of lower pre--SN radii, higher SN ejecta temperatures
and velocities and harder emission leads to SBO bolometric LCs with peaks similar
to those found for RSG and BSG progenitors ($10^{45}$-$10^{46}$~erg~s$^{-1}$).

The main, long-duration part of the LCs powered by
the radioactive decays of $^{56}$Ni and $^{56}$Co was studied for all models using the
{\it PHOENIX} results. For most PISN models, the implied $M_{\rm Ni}/M_{\rm ej}$
ratios were below 10~\%, a value close to Type IIP events leading to intrinsically
red colors, as noted by \citet{2012MNRAS.426L..76D}. As such, most PISN
exhibited peak luminosities spanning the range observed for normal
Type II and Ib/c events. Super-luminous LCs were found only in the most massive
non-zero metallicity 260~$M_{\odot}$ and 300~$M_{\odot}$ models.
We stress that super-luminous PISN LCs are possible for zero metallicity models with
higher masses than the ones we studied in this work ($>$~140~$M_{\odot}$).
Nevertheles, all models evolved with redder colors than those observed for SLSN-I and, more specifically, SN~2007bi.

We investigated the potential to distinguish rotation and metallicity information from
the observed PISN spectra and came to the conclusion that the spectroscopic
properties of these events are quite degenerate and any attempt to uncover
them based on observed spectra would be
highly uncertain. More specifically, we find that the increased abundances of
C, N and O in the outer layers of the rotating models do not have a significant
effect in the spectra and that, once the explosion sets-in the
products of nuclear burning are the same regardless of the pre--SN rotational mixing and metallicity.
The peak luminosity spectra, but also the spectral evolution of PISN in general,
appear to be quite similar across different initial metallicity and rotation rate.
This is in contradiction with the result of \citet{2011ApJ...734..102K} that increased
metallicity leads to suppressed near-UV flux and differences in some spectral features.
This difference is attributable to the fact that in their approach the extra metals were
added after the PISN explosion and prior to the free expansion phase where the radiative
transfer calculation is done. Self-consistent evolution of PISN with ZAMS metallicity
results in final spectra with similar properties.

Most of the spectroscopic properties across the PISN parameter space arise due to
fundamental differences in pre--SN composition and different SN ejecta masses, 
temperatures and velocities. The more massive models have a somewhat bluer continuum with
fewer lines in the optical than then less massive ones.
These points are well illustrated by our 90~$M_{\odot}$ model that was a pure
oxygen star prior to explosion. The oxygen-rich composition together with the
high SN ejecta temperature leads to spectra
with strong \ion{O}{1} absorption features that are distinguishable from those
of the other models. 

Our model PISN spectra compare well against those calculated 
by \citet{2011ApJ...734..102K} and \citet{2013MNRAS.428.3227D}
for similar progenitors (more specifically their bare 100~$M_{\odot}$ He core model He100).
This confirms that regardless of the differences in numerical methods between different codes and
LTE versus non--LTE treatment the results for the main emission characteristics of PISN
are consistent. 

Although the R-band LC of the 0.1~$Z_{\odot}$~300~$M_{\odot}$ model reproduces the main features
of the LC of SN~2007bi, it is not sufficient information to categorize the event as a PISN.
Comparison of our model spectra against the observed +54~d spectrum
of the super-luminous SN~2007bi at contemporaneous epochs confirms that
SN~2007bi was not a PISN, in agreement with the result of \citet{2013MNRAS.428.3227D}.
More specifically, PISNe appear to have redder color evolution and different spectroscopic
features compared to all observed SLSNe. 
This could be due to the fact that most
SLSNe have been observed in environments with metallicities $\sim$~0.45~$Z_{\odot}$
\citep{2014ApJ...787..138L} disfavoring the formation of very massive pre--SN CO cores
due to strong mass--loss. This value is near the upper limit metallicity of 1/3~$Z_{\odot}$
for the most massive progenitors assuming standard initial mass functions (IMF)
\citep{2007A&A...475L..19L}. In addition, other models of PISN present in the literature 
with LCs in agreement with that of SN~2007bi have adopted a progenitor 
metallicity that is far lower than that of the host galaxy of the event \citep{2014A&A...565A..70K} 
or have made simplifying assumptions with regards to the progenitor mass--loss history 
\citep{2008AIPC..990..263K, 2009Natur.462..624G}.
Given these results, we conclude that all SLSN events observed so far must be powered via alternative
mechanisms with the most promising candidates being radiation from a mangetar-spin down 
\citep{2012MNRAS.426L..76D, 2014MNRAS.437..656M} and CSM interaction of the H-rich and H-poor variety
\citep{2012ApJ...760..154C, 2013ApJ...773...76C, 2014MNRAS.441..289B, 2014arXiv1405.1325N}.

The PISN phenomenon is expected to be more prevalent in ultra-low and zero metallicity environments
appropriate to the conditions of the early Universe. Some Population III star-formation simulations predict
the first stars to be massive and rotating \citep{2012MNRAS.422..290S, 2013MNRAS.431.1470S},
well within the PISN parameter space. As such, the future of observational research in the subject
seems to rest on the shoulders of exciting upcoming missions such as {\it WFIRST} and {\it JWST}.

\acknowledgments

We would like to thank Wesley Even, 
Sean M. Couch, Donald Q. Lamb,
Ryan Chornock, Ken Chen, Alexandra Kozyreva 
and Ragnhild Lunnan for useful discussions and comments. 
This work was supported in part by the STScI grants AR 12820 and 13276 and the
National Science Foundation under grants
AST-0909132 and PHY-0822648 for the Physics Frontier Center ``Joint Institute for Nuclear Astrophysics" (JINA).
EC would like to thank the Enrico Fermi Institute for its support via the Enrico
Fermi Fellowship.  D.J.W. was supported by 
the European Research Council under the European Community's Seventh Framework Programme 
(FP7/2007 - 2013) via the ERC Advanced Grant "STARLIGHT:  Formation of the First Stars" (project 
number 339177).


\bibliography{pisnrefs}

\end{document}